\documentclass[
prd 
,showpacs, amssymb, superscriptaddress, aps
]{revtex4}
\usepackage{graphicx}
\input{epsf}

\usepackage{amsmath,amssymb}
\usepackage{bm}
\usepackage{times}

\newcommand{\dalm}{\kern1pt\vbox{\hrule height 0.9pt\hbox{\vrule width 0.9pt
\hskip 2.5pt\vbox{\vskip 5.5pt}\hskip 3pt\vrule width 0.3pt}\hrule height 0.3pt}
\kern1pt}

\newcommand{\be}{\begin{eqnarray}}
\newcommand{\ee}{\end{eqnarray}}
\newcommand{\beq}{\begin{eqnarray}}
\newcommand{\eeq}{\end{eqnarray}}

\usepackage{ulem}
\usepackage{color}

\begin{document}



\title{Pulse profiles of highly compact pulsars in general relativity}

\author{Hajime Sotani}
\email{sotani@yukawa.kyoto-u.ac.jp}
\affiliation{Division of Theoretical Astronomy, National Astronomical Observatory of Japan, 2-21-1 Osawa, Mitaka, Tokyo 181-8588, Japan}

\author{Umpei Miyamoto}
\affiliation{Research and Education Center for Comprehensive Science, Akita Prefectural University, Akita 015-0055, Japan}

\date{\today}

\begin{abstract}
Gravitational light bending by compact stars is an important astrophysical phenomenon. The bending angle depends on the stellar compactness, which is the ratio of stellar mass $M$ to radius $R$. In this paper, we investigate the pulse profile of highly compact rotating neutron stars for which the bending angle exceeds $\pi/2$. When $M/R > 0.284$ (the bending angle becomes equal to $\pi/2$ for the stellar model with $M/R=0.284$), such a large bending happens, resulting in that a photon emitted from any position on the stellar surface can reach an observer. First, we classify the parameter plane of inclination angle $i$ and angle $\Theta$ between the rotation axis and the normal on the hot spot by the number of photon paths reaching the observer. Then, we estimate the time-dependent flux of photons emitted from two hot spots on the rotating neutron star, associated with the magnetic polar caps, for various combinations of $i$ and $\Theta$, and for two values of compactness, assuming that the stellar rotation is not so fast that the frame dragging and the stellar deformation are negligible. As the result, we find that the pulse profiles of highly compact neutron stars are qualitatively different from those for the standard neutron stars. In particular, the ratio of the maximum observed flux to the minimum one is significantly larger than that for the standard neutron stars. This study suggests that one would be able to constrain the equation of state for neutron stars through the observation of pulse profile with angles $i$ and $\Theta$ determined by other methods.
\end{abstract}

%

\maketitle

\section{Introduction}
\label{sec:I}

Neutron stars are considered to be formed via the core-collapse supernovae. The density inside the neutron star significantly exceeds the nuclear saturation density $\sim 2.7\times 10^{14}$ g/cm$^3$, the surface magnetic fields may be quite strong $\sim10^{14}$ G (although the magnetic structure and strength are still uncertain), and also the gravitation is quite strong \cite{shapiro-teukolsky}. Since these extreme conditions are quite unique, neutron stars are thought of as suitable laboratories to probe the physics under such extreme conditions~\cite{Berti2015}. In fact, the discoveries of $2M_\odot$ neutron stars \cite{D2010,A2013} impose a severe restriction on the equation of state in high-density regime. Moreover, via the future observations of gravitational waves from a neutron star, one may be able to probe the gravitational theory in strong-field regime (see, e.g., \cite{SK2004,Sotani2014,Sotani2014a}).

The light bending by the gravity field around a star also can provide us the information of stellar structure. The strong lensing around the black hole is well-known as such an effect (e.g., \cite{VE2000,Bozza2002,SM2015}). In a similar way to the gravitational lensing around the black hole, the light ray emitted from neutron-star surface is bent due to the strong gravitational field produced by the neutron star. Therefore, one can expect that even the photon emitted from the backside of the neutron star may reach the observer, which significantly would contribute to the pulse profile from a rotating neutron star~\cite{PFC1983,LL95,Beloborodov2002}. Since the bending angle increases as the compactness, which is the ratio of the stellar mass to the radius, increases, the observation of pulse may enable us to obtain the information about the stellar compactness, which results in knowing the stellar mass and radius with the help of additional observation data~\cite{POC2014,Bogdanov2016}. On the other hand, via the pulse profiles of a rotating compact object, one may probe the gravitational geometry and gravitational theory \cite{SM2017,S2017}. In fact, such an attempt will be possible soon by the operation of x-ray timing mission with the Neutron star Interior Composition ExploreR (NICER)~\cite{NICER}.

To consider the pulse profiles of a rotating neutron star, if the spin frequency is not so high, one may be able to adopt the spherically symmetric solution as a background spacetime around the neutron star. So far, several groups have studied the pulse profiles using the Schwarzschild metric. 
In this context, Beloborodov derived the approximate formula of the observed flux coming from two polar caps on the pulsar and showed the classification how the photons radiated from two hot spots would be observed \cite{Beloborodov2002}. This approximation works very well for the neutron-star models with low compactness. In a similar fashion, we examined the pulse profiles in \cite{SM2017}, where we mainly focused on the sensitivity of light curve to the background spacetime and gravitational theories. Anyway, these past studies are only concerned with stellar models with an invisible zone, i.e., a part of neutron-star's surface that cannot be seen from a distant observe. Actually, the compactness of most neutron stars is not so high that the photon emitted from the their surface opposite to the observer is not bent enough.

The photon emitted from any position of neutron-star surface, however, can reach the observer, if the compactness is significantly high. Such a strong bending is possible, although the region in the mass-radius plane is limited (see Fig. \ref{fig:MR}), and the pulse profiles should be different from those for the standard neutron stars. Therefore, in this paper, we shall focus on the neutron stars with high compactness and discuss the pulse emanating from their two hot spots associated with the magnetic polar caps. 
As we will see, for the stellar models with high compactness, the classification of photon paths is  completely different from that obtained by Beloborodov \cite{Beloborodov2002}. Thus, first we will present such a classification for the neutron-star model with high compactness, then we will examine the pulse profiles for specific stellar models. As the result, it will be shown that the pulse profiles from the highly compact neutron stars are qualitatively different from those for the standard neutron stars.

In this paper, we adopt the geometrized units $c=G=1$, where $c$ and $G$ are the speed of light and the gravitational constant, respectively. The metric signature is $(-,+,+,+)$.

\section{Photon radiating from a hot spot}
\label{sec:II}

The equation of motion for a photon emitted from a hot spot on the neutron star has already been obtained in Ref.~\cite{SM2017}. Here, we just review it briefly. The metric of a static spherically symmetric spacetime is generally given by 
\begin{equation}
	g_{\mu\nu} dx^\mu dx^\nu
	= -A(r)dt^2 + B(r)dr^2 + C(r)\left(d\theta^2 + \sin^2\theta d\psi^2\right).   \label{eq:metric}
\end{equation}
We assume the asymptotic flatness as $A(r)\to 1$, $B(r)\to 1$, and $C(r)\to r^2$ as $r\to \infty$. In this paper, we shall particularly consider the Schwarzschild spacetime, for which the metric functions are given by
\begin{equation}
  A(r) = 1- \frac{2M}{r},\  B(r) = \frac{1}{A(r)},\   C(r) = r^2.
\end{equation}

\begin{figure}
\begin{center}
\includegraphics[scale=0.5]{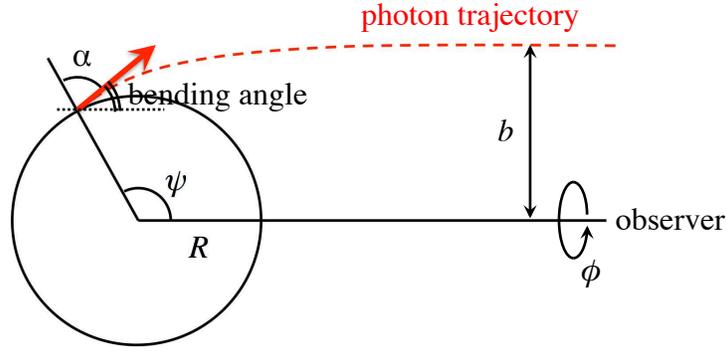} 
\end{center}
\caption{
Image of the photon trajectory radiating from the stellar surface. $R$ and $b$ denote the stellar radius and impact parameter, respectively, while $\alpha$ is the emission angle at the position of $\psi$.
}
\label{fig:trajectory}
\end{figure}

We consider a photon radiating from a small hot spot on the stellar surface, where the angle between the normal vector at the hot spot and the direction toward the observer is $\psi$, as shown in Fig. \ref{fig:trajectory}. Then, the angle at the stellar surface $r=R$ is given by
\begin{equation}
   \psi(R)
	= \int_R^\infty \frac{dr}{C}\left[\frac{1}{AB}\left(\frac{1}{b^2}-\frac{A}{C}\right)\right]^{-1/2},\label{eq:psi}
\end{equation}  
where $b$ is an impact parameter. Letting the emission angle be $\alpha$, which is the angle between the normal vector at the hot spot and the direction of the photon radiation from the hot spot, the impact parameter is given by
\begin{equation}
   b = \sin\alpha\sqrt{\frac{C(R)}{A(R)}}.  \label{eq:alpha}
\end{equation}

\begin{figure}
\begin{center}
\includegraphics[scale=0.5]{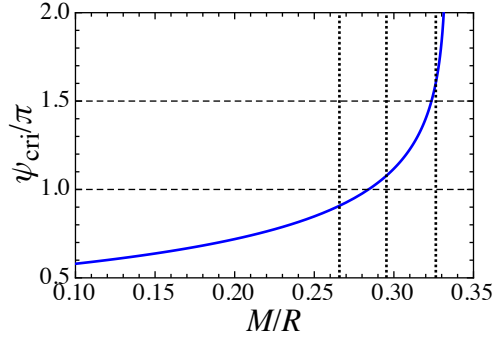} 
\end{center}
\caption{
The value of $\psi_{\rm cri}$ as a function of the stellar compactness $M/R$. The vertical dotted-lines correspond to the stellar models adopted in this study, whose compactness are respectively $M/R=0.2658$, $M/R=0.2953$, and $0.3263$ from left to right. These models respectively correspond to the neutron stars with $M=1.8M_\odot$, $2.0M_\odot$, and $2.21M_\odot$, fixing $R=10$ km, which are marked with the circle, pulse, and cross in Fig.~\ref{fig:MR}.
}
\label{fig:com-psi}
\end{figure}

One can numerically obtain the relation between $\psi(R)$ and $\alpha$ for given $R$ via Eqs.~(\ref{eq:psi}) and (\ref{eq:alpha}). The difference of angles $\psi(R)-\alpha$, which is a bending angle, does not vanish due to the light bending by the gravitation of neutron stars. Angle $\psi(R)$ increases as $\alpha$ increases, and reaches a maximum at $\alpha=\pi/2$. In particular, the maximum value of $\psi(R)$ is denoted as $\psi_{\rm cri}$ and depends on the compactness of neutron star. 
In Fig.~\ref{fig:com-psi}, the value of $\psi_{\rm cri}$ is shown as a function of the stellar compactness $M/R$. From this figure, one can see that $\psi_{\rm cri}$ becomes larger than $\pi$ for the stellar model with high compactness. In such a stellar model, the invisible zone on the stellar surface disappears and the photons radiated from any position on the stellar surface can reach the observer. 
In this paper, we particularly focus on the neutron-star models for which $\psi_{\rm cri}>\pi$, and examine the pulse profiles from such neutron stars. In the case of $\psi_{\rm cri}>\pi$, the number of photon paths can be multiple, i.e., the photon emitted from the stellar surface can proceeds in the directions of the decrease and increase of $\psi$ to reach the distant observer .

The observed flux $dF$ of the photons radiated from the hot spot whose area is $dS$ is given by
\begin{equation}
   dF = I_0(\alpha) A(R) \cos\alpha \frac{d(\cos\alpha)}{d\mu} \frac{dS}{D^2}, 
   \label{eq:dF0}
\end{equation}
where $\mu := \cos\psi$, $I_0$ is the surface intensity, and $D$ is the distance between the observer and the star~\cite{Beloborodov2002,SM2017}. 
Setting $\phi$ as an azimuthal angle with respect to the direction to the observer from the stellar center, $dS$ in Eq. (\ref{eq:dF0}) becomes $dS=R^2\sin\psi d\psi d\phi$. Then, Eq. (\ref{eq:dF0}) can be rewritten as
\begin{equation}
  dF =  I_0(\alpha) A(R) \sin\alpha \cos\alpha \frac{d\alpha}{d\psi} \frac{R^2}{D^2} d\psi d\phi.
\end{equation}
As in \cite{Beloborodov2002,SM2017}, adopting the pointlike-spot approximation for simplicity, where the spot area is assumed to be so small that the variables in Eq. (\ref{eq:dF0}) do not depend on the position in the area of $dS$, and integrating $dF$ in the ranges of $\psi-\delta\psi\le\psi\le\psi+\delta\psi$ and $\phi-\delta\phi\le\phi\le\phi+\delta\phi$, one obtains the observed bolometric flux as
\begin{equation}
   F_*(\psi) :=\int dF =  I_0(\alpha) \frac{4A(R)R^2\delta\psi \delta\phi}{D^2}\sin\alpha\cos\alpha\frac{d\alpha}{d\psi}.
\end{equation}
Although $I_0$ generally depends on the emission angle $\alpha$, hereafter we assume the isotropic emission in a local Lorentz frame, i.e., $I_0 = {\rm const.}$ for simplicity. Then, the observed flux is given by 
\begin{equation}
   F_*(\psi) = F_0 \sin\alpha\cos\alpha\frac{d\alpha}{d\psi},  
\;\;\;
	F_0 : = \frac{4I_0 A(R)R^2\delta\psi \delta\phi}{D^2}.
\label{eq:dF1}
\end{equation}
We remark that this observed flux comes from the spot whose area is $S_0:=\int dS=4R^2\delta\psi\delta\phi\sin\psi$, while the area of $S_0$ depends on the position $\psi$. Now, considering the observed flux from the hot spot on the stellar surface, whose area is fixed to be $s$, the observed flux $F(\psi)$ is given by $F(\psi) = F_*(\psi)\times s/S_0$. Finally, the observed flux is expressed as
\begin{equation}
	  F(\psi) = F_1 \cos\alpha\frac{d(\cos\alpha)}{d\mu},  
\;\;\;
	F_1 : = I_0 \frac{ s A(R)}{D^2},
\label{eq:dF2}
\end{equation}
which is the same expression in \cite{Beloborodov2002,SM2017}.
Due to the coordinate singularity of the polar coordinates at poles $\psi=0$ and $\pi$ (i.e., the points where $d\mu = -\sin \psi d\psi$ appearing in Eq.~\eqref{eq:dF2} vanishes), one cannot treat the flux from the hot spot when it is exactly on either pole by the present scheme, as long as we rely on the point-like approximation of hot spot. Although such a situation occurs only when $i=\Theta$ holds, we do not consider such a special situation in this paper, where the meaning of the angles $i$ and $\Theta$ is as shown in Fig. \ref{fig:pulsar}.

\section{Adopted neutron-star models}
\label{sec:III}

\begin{figure}
\begin{center}
\includegraphics[scale=0.5]{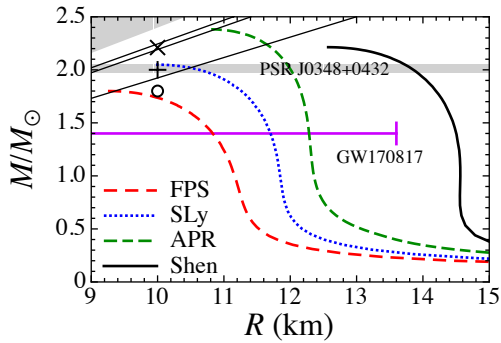} 
\end{center}
\caption{
Stellar models adopted in this study are shown with the circle, plus, and cross. The plus symbol denotes the neutron-star model with $M=2.0M_\odot$ and $R=10$ km, for which $\psi_{\rm cri}=1.078\pi$, i.e., $\pi<\psi_{\rm cri}<3\pi/2$, while the cross symbol denotes that with $M=2.21M_\odot$ and $R=10$ km, for which $\psi_{\rm cri}=1.604\pi$, i.e., $3\pi/2<\psi_{\rm cri}<2\pi$. For reference, the neutron-star model with $M=1.8M_\odot$ and $R=10$ km, for which $\psi_{\rm cri}=0.908\pi$, i.e., $\psi_{\rm cri}<\pi$, is also marked with the circle. For reference, mass and radius relations constructed with several EOSs are shown. Three slid-straight lines denote the stellar models with $\psi_{\rm cri}=\pi$, $3\pi/2$, and $2\pi$, where $M/R=0.2840$, $0.3236$, and $0.3313$, from bottom to top. The top-left painted region is forbidden by the causality~\cite{L2012}. In addition, the observed maximum mass of PSR J0348+0432 is shown with the horizontal thick-solid line and the radius constraint with the gravitational wave observation (GW170817) is shown by the horizontal solid line.
}
\label{fig:MR}
\end{figure}

Here, let us introduce the neutron-star models adopted in this study. Up to now, many equations of state (EOSs) for neutron-star matter have been proposed theoretically, 
but it is not fixed yet.
This is because the density inside the star significantly exceeds the nuclear saturation density, while the constraint on the properties of nuclear matter in such a high density region is quite difficult by terrestrial nuclear experiments.
Even so, recent astronomical observations set constraints on the EOS. One of them is the discoveries of neutron stars with $\sim 2M_\odot$, i.e., PSR J1614-2230 with $M=(1.97\pm0.04)M_\odot$ \cite{D2010} and PSR J0348+0432 with $M = (2.01\pm0.04)/M_\odot$ \cite{A2013}. Owing to the existence of such massive neutron stars, the EOS which predicts that the maximum mass is less than $2M_\odot$ is ruled out. Meanwhile, the observation of gravitational wave from the binary neutron-star merger \cite{GW6}, GW170817, gives us the constraint on the tidal deformability, which leads to another constraint on the stellar radius, i.e., the maximum radius of the neutron star with $1.4M_\odot$ is 13.6 km \cite{RM14}. In Fig.~\ref{fig:MR}, for reference, we plot the mass-radius relations constructed with several realistic EOSs, where the observed maximum mass of PSR J0348+0432 (horizontal thick-solid line) and the constraint on the $1.4M_\odot$ neutron-star radius (horizontal solid line) are also shown. 
We remark that FPS and SLy EOSs are based on the Skyrme-type effective interactions \cite{FPS,SLy}, Shen EOS is based on the relativistic mean fled theory \cite{Shen}, and APR EOS is with variational method \cite{APR}.
From this figure, FPS is ruled out from the $2M_\odot$ neutron-star observations, while Shen is ruled out from the radius constraint with GW170817. In addition to the observational constraints, the causality gives us theoretical constraint, i.e, $R < 2.824 M$~\cite{L2012}, which is shown by the top-left painted region.

For considering the pulse profile from a rotating neutron star, the stellar compactness is the most important parameter. In fact, under the assumption that the spin frequency is not so high, the pulse profile from a neutron star is the same as that from another neutron star with the same compactness, even if the mass and radius are different from each other \cite{Beloborodov2002,SM2017}. Moreover, as mentioned in the previous section, since we focus on the neutron-star models for which $\psi_{\rm cri}>\pi$, in Fig. \ref{fig:MR} we plot the solid-straight lines corresponding to the stellar models with $\psi_{\rm cri}=\pi$, $3\pi/2$, and $2\pi$ from bottom to top, i.e,  $M/R=0.2840$, $0.3236$, and $0.3313$ from bottom to top.  From this figure, one sees that $\psi_{\rm cri} < \pi$ holds for most neutron-star models. However, one can see also that the stellar models with $\psi_{\rm cri} >\pi$ are still allowed although such a region is limited in the mass-radius plane.

In order to examine the pulse profiles for the neutron-star models whose values of $\psi_{\rm cri}$ are in the range of $\pi< \psi_{\rm cri}<3\pi/2$ and $3\pi/2< \psi_{\rm cri}<2\pi$, in this study we particularly adopt two neutron-star models with $(M, R)=(2.0M_\odot, 10 {\rm km})$ and $(2.21M_\odot, 10 {\rm km})$, which are shown in Fig.~\ref{fig:MR} with the plus and cross symbols, respectively. Additionally, for reference we consider the neutron-star model with $(M, R)=(1.8M_\odot, 10 {\rm km})$ in the appendix, which is also marked with the circle in Fig.~\ref{fig:MR}. In fact, since the pulse profile depends on only the stellar compactness, the stellar models marked with the plus and circle can be constructed with SLy or APR EOSs with different mass and radius, while the stellar model marked with the cross seems to correspond to the stellar model with the maximum mass constructed with APR EOS with different mass and radius.

\section{Classification of Pulse profiles}
\label{sec:IV}

\begin{figure}
\begin{center}
\includegraphics[scale=0.4]{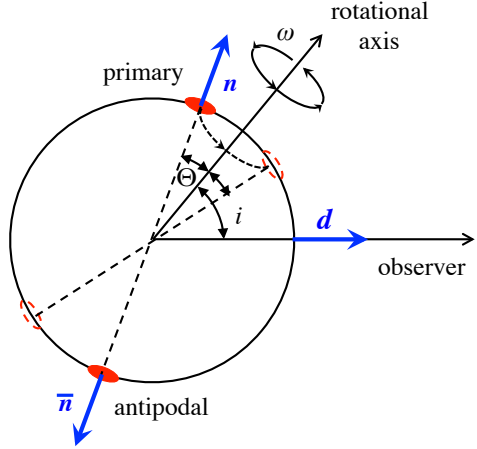} 
\end{center}
\caption{
Schematic picture of the hot spots on the pulsar rotating with angular velocity $\omega$. $\bm{d}$ is the unit vector pointing to the observer, while $\bm{n}$ and $\bar{\bm {n}}$ are respectively the normal unit vectors on the primary and antipodal hot spots. $\Theta$ is the angle between the rotational axis and $\bm{n}$, while $i$ is the angle between the rotational axis and $\bm{d}$. Angles $i$ and $\Theta$ are in the range of $0\le i\le\pi/2$ and $0\le \Theta \le \pi/2$.
}
\label{fig:pulsar}
\end{figure}

As in Refs.~\cite{Beloborodov2002,SM2017}, we consider a rotating neutron star with two antipodal hot spots associated with the magnetic polar caps. As shown in Fig.~\ref{fig:pulsar}, we call the hot spot closer to the observer a primary one and the other an antipodal one. The unit vector pointing the observer is denoted by $\bm{d}$, while the unit normal vectors on the primary and antipodal hot spots are respectively denoted by $\bm{n}$ and $\bar{\bm{n}}$. The angle between $\bm{d}$ and the rotational axis is denoted by $i$, while the angle between $\bm{n}$ and the rotational axis is by $\Theta$. In this case, the angles $i$ and $\Theta$ are in the range of $0\le i\le \pi/2$ and $0\le\Theta\le \pi/2$. 

Regarding the angle between $\bm{n}$ and $\bm{d}$ as $\psi$ for the pulsar rotating with angular velocity $\omega$, one obtains the time dependence of $\mu=\cos\psi$ for given $i$ and $\Theta$ as
\begin{equation}
   \mu(t)=\sin i\sin\Theta\cos(\omega t) + \cos i\cos\Theta,  \label{eq:mut_p}
\end{equation}
where we set $t=0$ when the primary hot spot is closest to the observer, i.e., when $\mu$ has a maximum value. We note that, at any given point in time, one can suppose the existence of a plane spanned by two vectors $\bm{d}$ and $\bm{n}$ (or $\bar{\bm{n}}$), in which the stellar center exists, and that the observer detects the photon from the hot spot whose trajectory is on such a plane. 

From Fig.~\ref{fig:pulsar}, the minimum and maximum values of $\mu$ are determined by
\begin{gather}
  \mu_{\rm min} = \cos\psi_{\rm max} = \cos(i+\Theta), \\
  \mu_{\rm max} = \cos\psi_{\rm min} = \cos|i-\Theta|,
\end{gather}
where $\psi_{\rm max}$ and $\psi_{\rm min}$ respectively denote the angles between $\bm{d}$ and $\bm{n}$ when the primary hot spot is closest to and farthest from the observer. On the other hand, the value of $\bar{\mu}$, which is for the antipodal hot spot, is given by $\bar\mu=-\mu$ since $\bm{\bar{n}}=-\bm{n}$. Thus, the minimum and maximum values of $\bar\mu$ are given by
\begin{gather}
  \bar\mu_{\rm min} = -\mu_{\rm max} = -\cos|i-\Theta|, \\
  \bar\mu_{\rm max} = -\mu_{\rm min} = -\cos(i+\Theta).
\end{gather}
Note that $\mu_{\rm max}>0$ and $\bar\mu_{\rm min}<0$ by definition.

As mentioned before, we consider the case of $\psi_{\rm cri}>\pi$. Therefore, the observer can always see the both hot spots. The number of photon paths, however, can be multiple, depending on $i$ and $\Theta$. 
For example, considering the stellar model with $\pi< \psi_{\rm cri}<3\pi/2$, the primary hot spot has only one clockwise photon path in Fig.~\ref{fig:pulsar} if $\mu_{\rm min}>\cos\psi_{\rm cri}$, while that spot has two photon paths, which are in the clockwise and counterclockwise directions, if $\mu_{\rm min}<\cos\psi_{\rm cri}$. 
That is, the photon trajectory from the primary hot spot may be not only clockwise but also counterclockwise, depending on $i$ and $\Theta$. We shall separately discuss the cases of $\pi< \psi_{\rm cri}<3\pi/2$ in \S \ref{sec:IVa} and $3\pi/2< \psi_{\rm cri}<2\pi$ in \S \ref{sec:IVb}. In these cases, the number of photon paths from each hot spot is either one or two. For reference, the classification for $\psi_{\rm cri}<\pi$ is presented in Appendix~\ref{sec:a1}. Note that the number of photon paths is possible to be more than two if one considers for $\psi_{\rm cri}>2\pi$, although the corresponding neutron-star models are extremely limited. 
For example, considering the stellar model with $2\pi<\psi_{\rm cri}<3\pi$, in addition to the photon paths in the clockwise and counterclockwise directions as mentioned the above, the photon path can exist where the photon reaches the observer in the clockwise path after making a circuit around the star. That is, in this case the photon paths can be maximally three. In a similar way, 
the maximum number of photon paths is $j \in {\bm N}$ in the case of $(j-1)\pi<\psi_{\rm cri}<j\pi$.

\subsection{Case of $\pi< \psi_{\rm cri}<3\pi/2$}
\label{sec:IVa}

Since $\mu_{\rm max}>0$, $\mu_{\rm max}$ is always larger than $\cos\psi_{\rm cri}$. Therefore, regarding the number of photon paths from the primary hot spot, one can consider the following two cases.
\begin{itemize}
  \item[{p$_1$}:]
    $\mu_{\rm min}>\cos\psi_{\rm cri}$, where the primary hot spot always has only one photon path.
  \item[{p$_2$}:]
    $\mu_{\rm min}<\cos\psi_{\rm cri}$, where the primary hot spot sometime has two photon paths.    
\end{itemize}
On the other hand, regarding the number of photon paths from the antipodal hot spot, the following three cases are possible.
\begin{itemize}
  \item[{a$_1$}:]
    $\bar\mu_{\rm max}<\cos\psi_{\rm cri}$, where the antipodal hot spot always has two photon paths.
  \item[{a$_2$}:]
    $\bar\mu_{\rm min}>\cos\psi_{\rm cri}$, where the antipodal hot spot always has only one photon path.    
  \item[{a$_3$}:]
    $\bar\mu_{\rm min}<\cos\psi_{\rm cri}<\bar\mu_{\rm max}$, where the antipodal hot spot sometime has two photon paths.
\end{itemize}

\begin{figure}
\begin{center}
\includegraphics[scale=0.5]{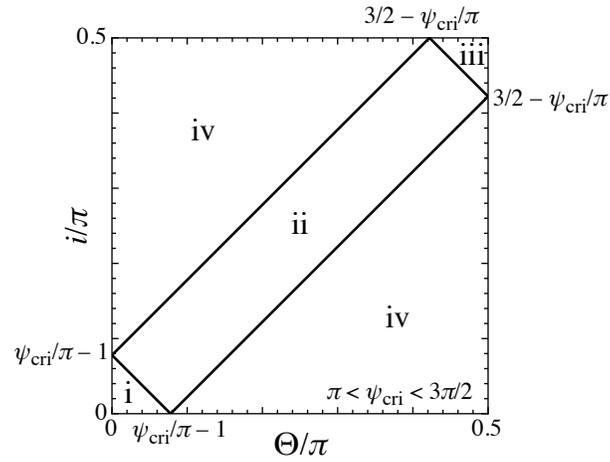} 
\end{center}
\caption{
For $\pi< \psi_{\rm cri}<3\pi/2$, the classification of the number of photon paths from the primary and antipodal hot spots is shown in the plane of angles $i$ and $\Theta$ for the neutron-star model with $M=2.0M_\odot$ and $R=10$ km (the plus symbol in Fig.~\ref{fig:MR}), for which $\psi_{\rm cri}=1.078\pi$.
}
\label{fig:theta-i-a0}
\end{figure}

Since the observed flux from a rotating neutron star is the superposition of flux radiated from both hot spots, there are the following four cases,
\begin{itemize}
  \item[{(i)}]
    Combination of p$_1$ and a$_1$.
  \item[{(ii)}]
    Combination of p$_1$ and a$_3$.
 \item[{(iii)}]
    Combination of p$_2$ and a$_3$.
  \item[{(iv)}]
    Combination of p$_1$ and a$_2$.
\end{itemize}
One can write down the condition corresponding to the each case in terms of $i$ and $\Theta$. The division of $(i, \Theta)$ plane according to the above classification is presented in Fig.~\ref{fig:theta-i-a0} for the neutron-star model with $M=2.0M_\odot$ and $R=10$ km.


\subsection{Case of $3\pi/2< \psi_{\rm cri}<2\pi$}
\label{sec:IVb}

\begin{figure}
\begin{center}
\includegraphics[scale=0.5]{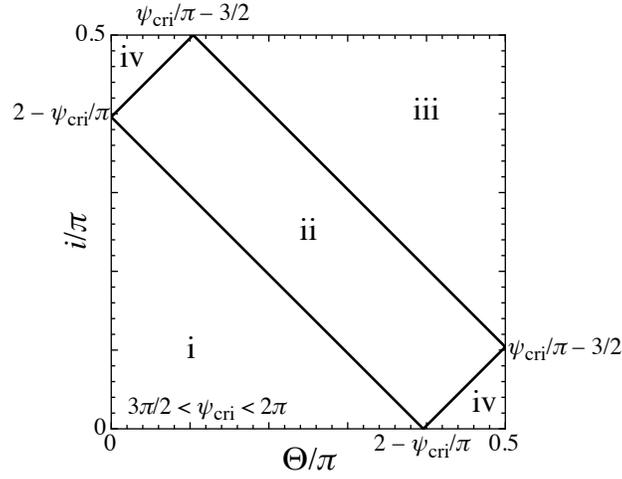} 
\end{center}
\caption{
For $\pi< \psi_{\rm cri}<3\pi/2$, the classification of the number of photon paths from the primary and antipodal hot spots is depicted in the plane of angles $i$ and $\Theta$ for the neutron-star model with $M=2.21M_\odot$ and $R=10$ km (the cross symbol in Fig.~\ref{fig:MR}), where $\psi_{\rm cri}=1.604\pi$. 
}
\label{fig:theta-i-b0}
\end{figure}

In this case, regarding the photon paths from the primary hot spot, there are following three cases.
\begin{itemize}
  \item[{p$_1$}:]
    $\mu_{\rm min}>\cos\psi_{\rm cri}$, where the primary hot spot always has only one photon path.
  \item[{p$_2$}:]
    $\mu_{\rm max}<\cos\psi_{\rm cri}$, where the primary hot spot always has two photon paths.    
  \item[{p$_3$}:]
    $\mu_{\rm min}<\cos\psi_{\rm cri}<\mu_{\rm max}$, where the primary hot spot sometime has two photon paths.
\end{itemize}
On the other hand, because $\bar\mu_{\rm min}<0$ (i.e., $\bar\mu_{\rm min}$ is always smaller than $\cos\psi_{\rm cri}$), regarding the number of photon paths from the antipodal hot spot, there are following two cases.
\begin{itemize}
  \item[{a$_1$}:]
    $\bar\mu_{\rm max}<\cos\psi_{\rm cri}$, where the antipodal hot spot always has two photon paths.
  \item[{a$_2$}:]
    $\bar\mu_{\rm max}>\cos\psi_{\rm cri}$, where the antipodal hot spot sometime has two photon paths.    
\end{itemize}
Taking into account all the above classification, there are four cases regarding the number of photon paths reaching the observer as Fig.~\ref{fig:theta-i-b0}, which is drawn for the neutron-star model with $M=2.21M_\odot$ and $R=10$ km. The regions denoted by i, ii, iii, and iv correspond to the followings.
\begin{itemize}
  \item[{(i)}]
    Combination of p$_1$ and a$_1$.
  \item[{(ii)}]
    Combination of p$_3$ and a$_1$.
 \item[{(iii)}]
    Combination of p$_3$ and a$_2$.
  \item[{(iv)}]
    Combination of p$_2$ and a$_1$.
\end{itemize}


\section{Pulse profiles}
\label{sec:V}

At any given point in time, $\psi(t)$, which represents the position of primary hot spot, is determined by Eq.~(\ref{eq:mut_p}) for given angles $i$ and $\Theta$. Using such a value of $\psi$, the flux from the primary hot spot is calculated with Eq.~(\ref{eq:dF2}), while that from the antipodal hot spot $\bar{F}$ is with Eq.~(\ref{eq:dF2}) by replacing $\psi$ by $\psi+\pi$. Then, the observed flux $F_{\rm ob}$ is determined by $F_{\rm ob}=F+\bar{F}$. In order to see the pulse profiles from the rotating neutron star, we adopt two neutron-star models with $(M, R)=(2.0M_\odot, 10 {\rm km})$ and $(2.21M_\odot, 10 {\rm km})$, which are shown in Fig.~\ref{fig:MR} with the plus and cross symbols, respectively. In order to compare these results with the case of $\psi_{\rm cri}<\pi$, we also consider the neutron-star model with $M=1.8M_\odot$ and $R=10$ km (the circle in Fig.~\ref{fig:MR}) in Appendix~\ref{sec:a1}.

\begin{figure*}
\begin{center}
\begin{tabular}{cc}
\includegraphics[scale=0.5]{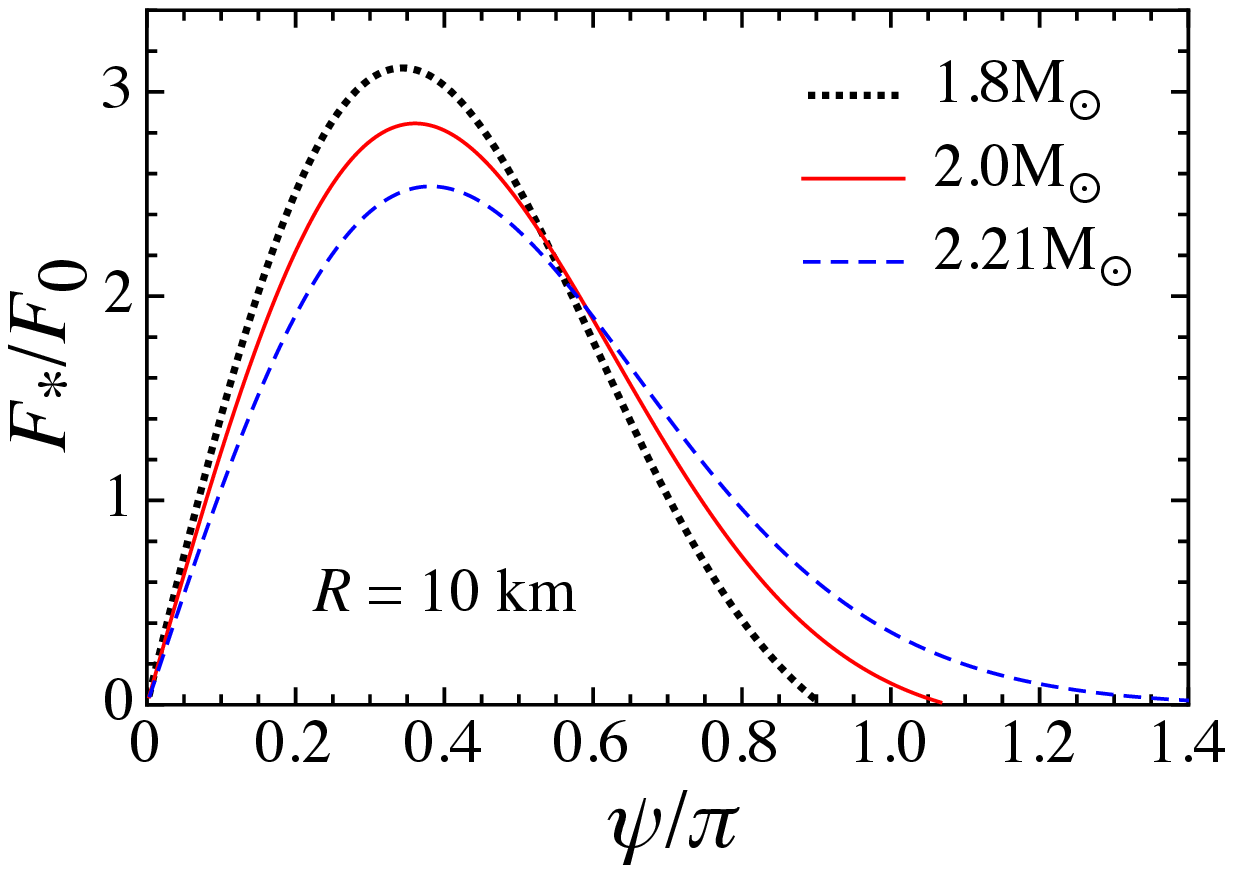} &
\includegraphics[scale=0.5]{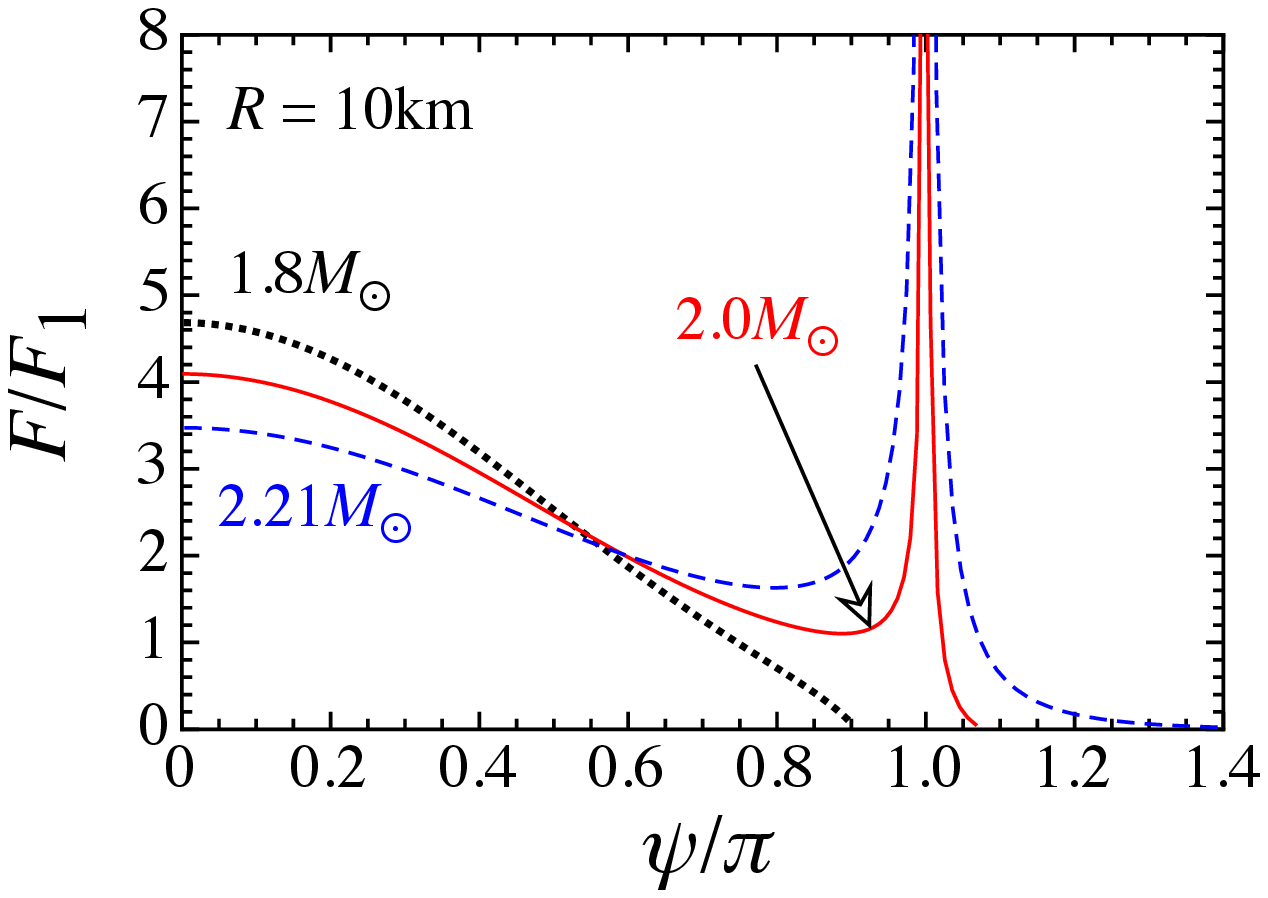} 
\end{tabular}
\end{center}
\caption{
$F_*/F_0$ given by Eq. (\ref{eq:dF1}) and $F/F_1$ given by Eq. (\ref{eq:dF2}) are shown as a function of $\psi/\pi$ for the neutron star models with $M=1.8M_\odot$ (dotted line), $2.0M_\odot$ (solid line), and $2.21M_\odot$ (dashed line), where the radius is fixed to be 10 km.
}
\label{fig:Fpsi}
\end{figure*}

Since $F_1$ in Eq.~(\ref{eq:dF2}) is independent of time, the pulse profile is determined by $F/F_1$, as $\mu(t)$ varies with time. So, before showing the pulse profiles for specific values of $i$ and $\Theta$, we examine the behavior of $F/F_1$. In Fig.~\ref{fig:Fpsi}, 
$F_*/F_0$ and $F/F_1$ are plotted as functions of $\psi/\pi$ for the neutron-star models with $M=1.8M_\odot$ by dotted line, $2.0M_\odot$ by solid line, and $2.21M_\odot$ by dashed line, fixing the radius to $R=10$ km.

From Fig.~\ref{fig:Fpsi}, one can see that the flux $F_*/F_0$ smoothly changes as $\psi$ increases for any stellar models. 
On the other hand, for $\pi<\psi_{\rm cri}<2\pi$, the dependence of $F/F_1$ on $\psi$ is obviously different from that for $\psi_{\rm cri}<\pi$, i.e., as $\psi$ increases, $F/F_1$ decreases first but then begins increasing before $\psi = \pi$. If the hot spot comes close to $\psi\sim\pi$, i.e., $|i-\Theta|\sim 0$, the flux from such a hot spot becomes important for pulse profiles. 
This brightening around $\psi = \pi$ seems to be the result of that we consider the flux from the hot spot whose area is fixed. In other words, in order to keep the spot area, the value of $\delta\psi\delta\phi$ in Eq.~(\ref{eq:dF1}) depends on the position of the spot. Then, if the spot would approach the position of $\psi \sim \pi$, the value of $\delta\psi\delta\phi$ increases. In fact, the photon path we consider here is confined on the plane spanned by $\bm{n}$ and $\bm{d}$ in Fig.\ \ref{fig:pulsar}, while if the spot comes to the position of $\psi=\pi$, which cannot be treated in our formalism, the number of the photon paths becomes infinity because one cannot choose a specific plane. Anyway, we note that the behavior in the vicinity of $\psi=\pi$ may be modified and one could remove the singularity at $\psi=\pi$ if the pointlike-spot approximation is not applied. In addition, we find that $F/F_1$ is quite small for $\psi>\pi$ compared to that for $\psi<\pi$. That is, if the angle $\psi$ is larger than $\pi$, the flux from such a hot spot does not contribute significantly to the pulse profile.



Here, let us give a few comments on the behavior of $F/F_1$. Firstly, although the flux for $\psi_{\rm cri}>\pi$ seems to diverge at $\psi/\pi=1$, it should not be taken as it is. Namely, as mentioned at the end of \S \ref{sec:II}, the flux exactly from the coordinate poles ($\psi = 0$ and $\pi$) cannot be dealt appropriately in the present framework, relying on the point-like approximation of hot spot. Secondly, it can be analytically proven that $F/F_1$ vanishes at $\psi=\psi_{\rm cri}$ by expanding Eq.~\eqref{eq:dF2} around $\psi_{\rm cri}$ and substituting $\alpha(\psi_{\rm cri})=\pi/2$.

\begin{figure*}
\begin{center}
\begin{tabular}{cc}
\includegraphics[scale=0.5]{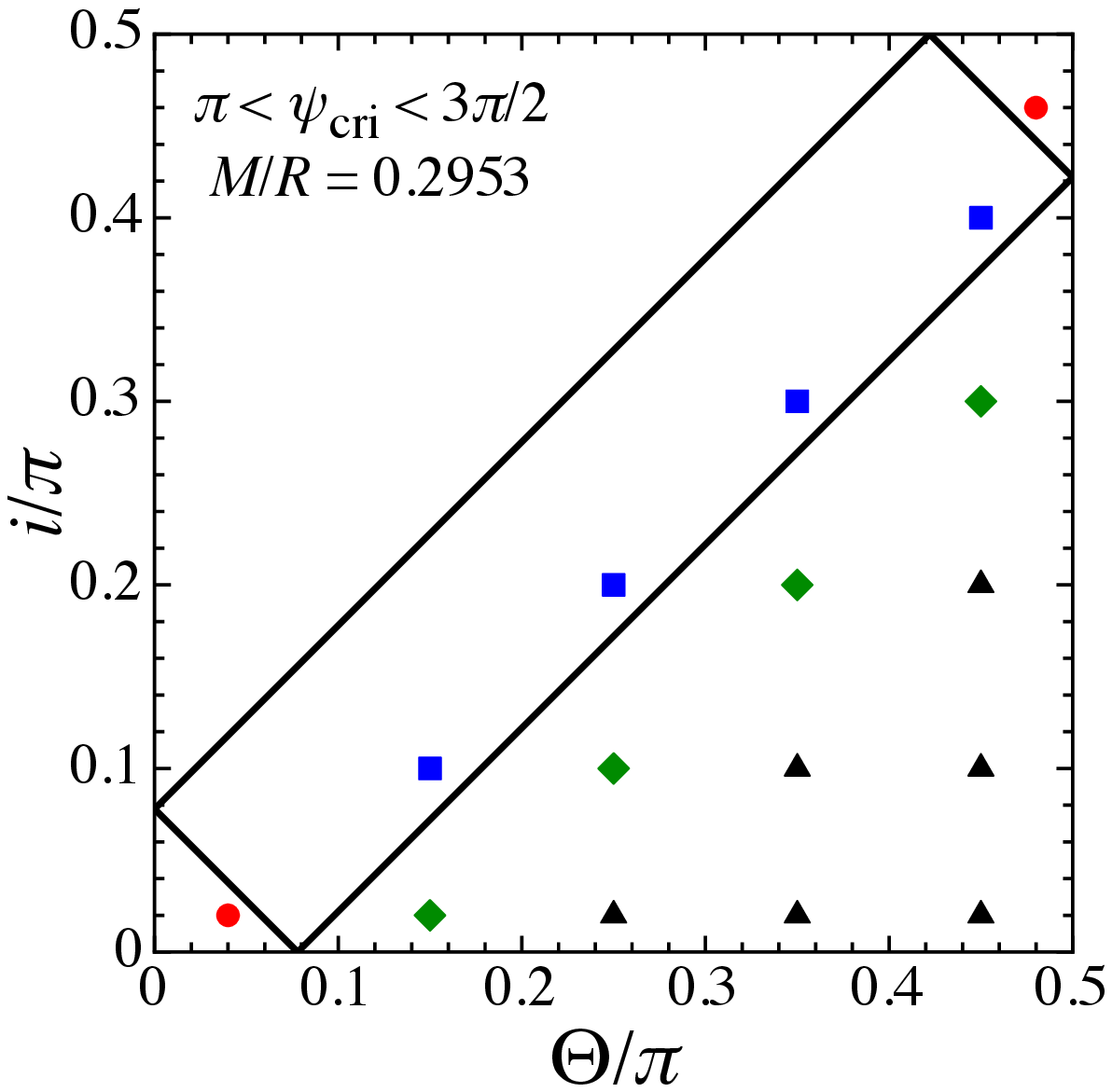} &
\includegraphics[scale=0.5]{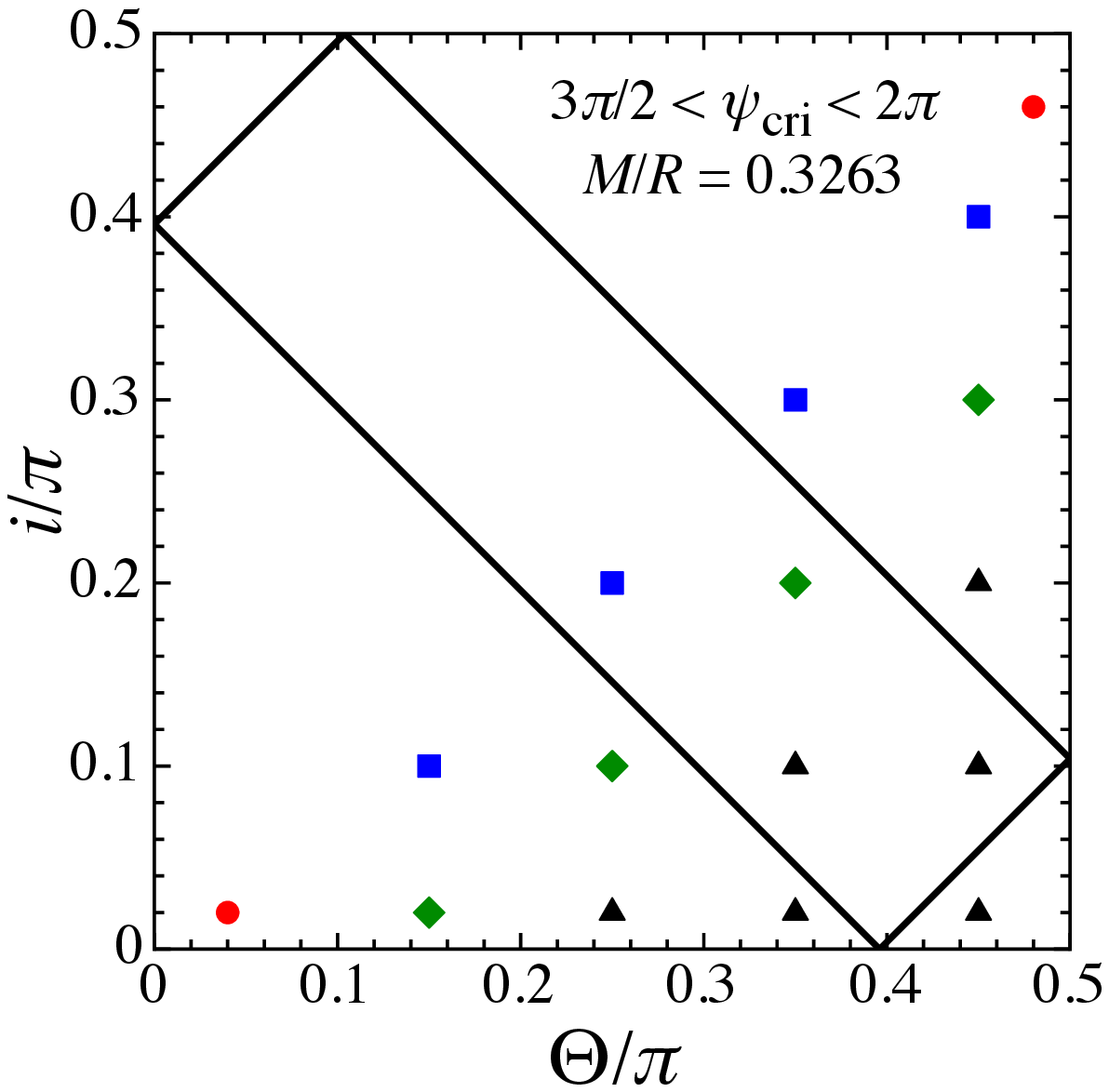} 
\end{tabular}
\end{center}
\caption{
Specific angles of $i$ and $\Theta$, with which the pulse profiles are considered in this paper, on the classification for the stellar models with $M/R=0.2953$ (left panel) and 0.3263 (right panel). In the figure, the circles, squares, diamonds, and triangles correspond to the stellar models for the pulse profiles shown in Figs.~\ref{fig:pulse-M200} and \ref{fig:pulse-M221} from left to right panels. 
}
\label{fig:itheta}
\end{figure*}

\begin{figure*}
\begin{center}
\begin{tabular}{c}
\includegraphics[scale=0.38]{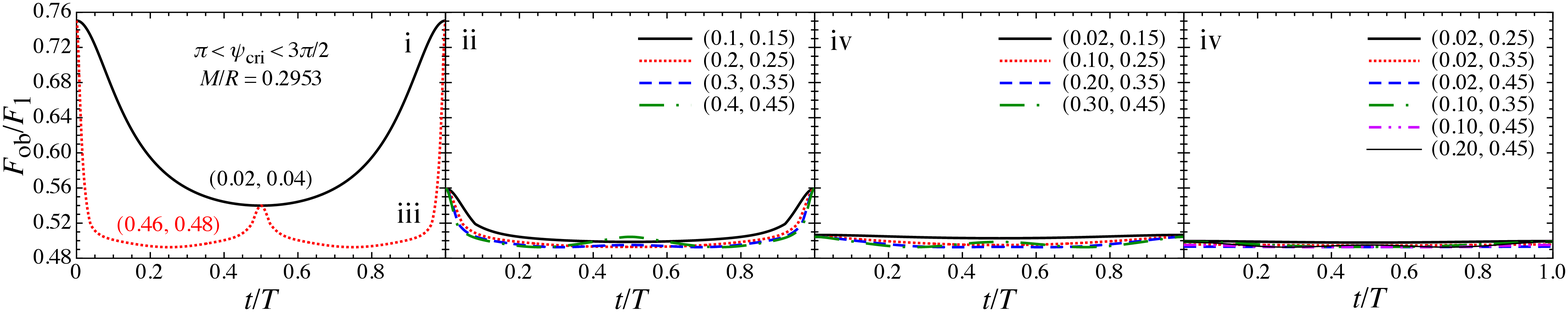} \\
\includegraphics[scale=0.38]{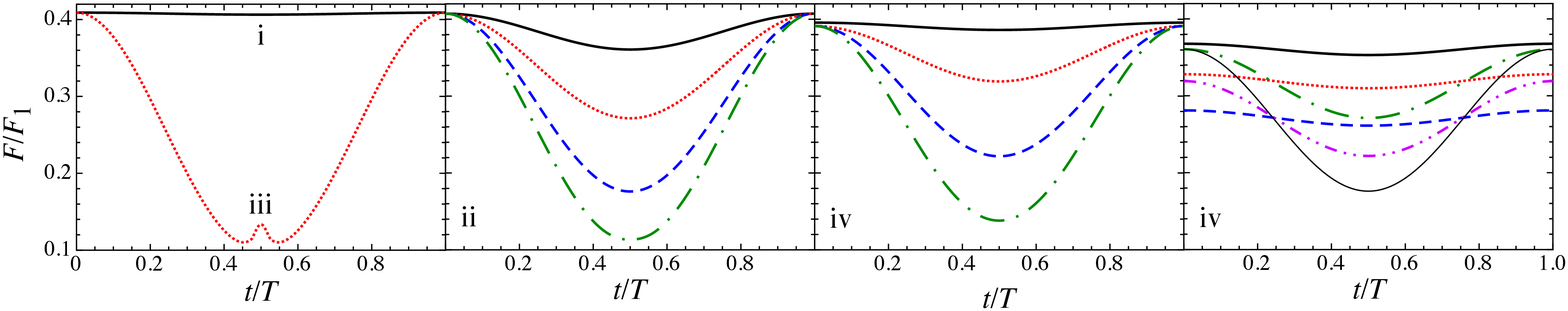} \\ 
\includegraphics[scale=0.38]{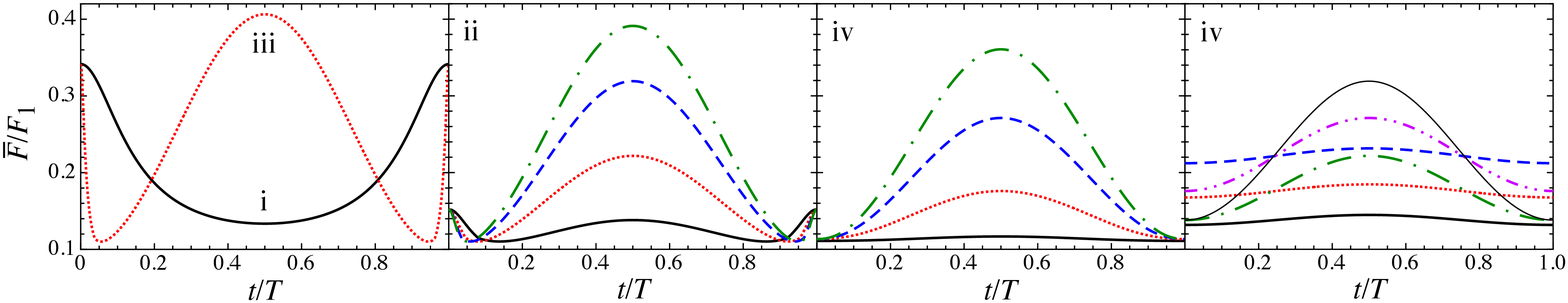}  
\end{tabular}
\end{center}
\caption{
Pulse profiles for the neutron star model with $M=2.0M_\odot$ and $=10$ km are shown as a function of $t/T$, where $T:=2\pi/\omega$, for various angles $i$ and $\Theta$ shown in Fig.~\ref{fig:itheta}. The panels from left to right correspond to the angles shown in Fig.~\ref{fig:itheta} with the circles, squares, diamonds, and triangles. The panels from top to bottom are the observed flux $F_{\rm ob}$, the flux from the primary hot spot $F$, and that from the antipodal hot spot $\bar{F}$, normalized by $F_1$. The labels of i, ii, iii, and iv denote the classes of pulse profiles depending on the angles $i$ and $\Theta$, which are explained in the text. 
}
\label{fig:pulse-M200}
\end{figure*}

\begin{figure*}
\begin{center}
\begin{tabular}{c}
\includegraphics[scale=0.38]{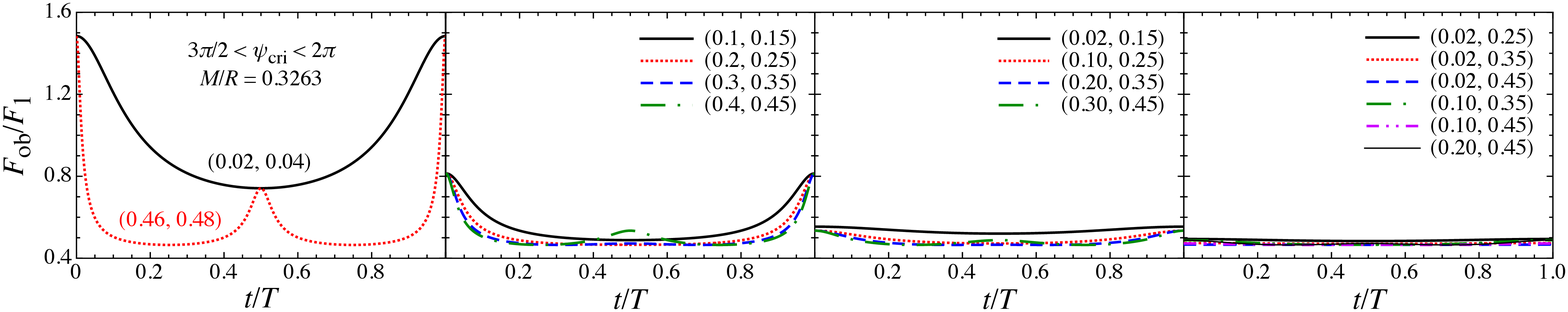} \\
\includegraphics[scale=0.38]{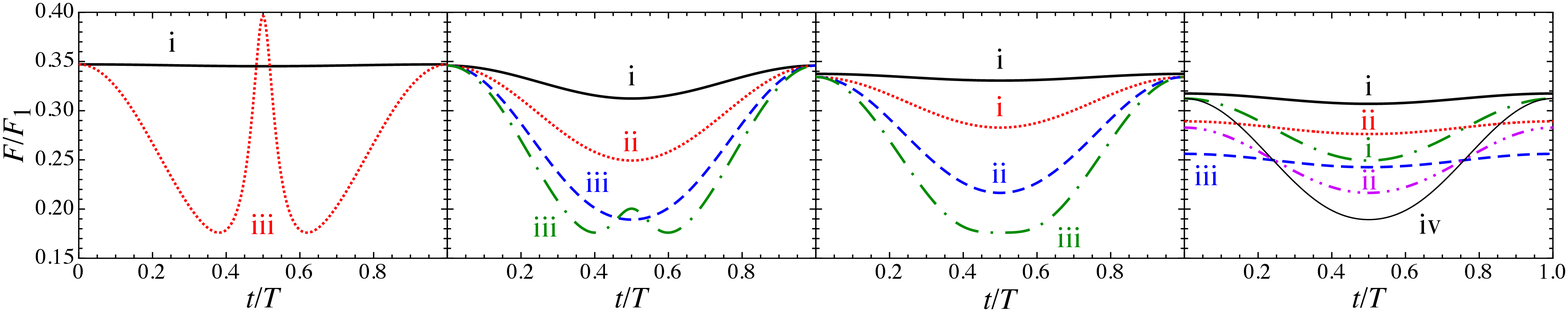} \\ 
\includegraphics[scale=0.38]{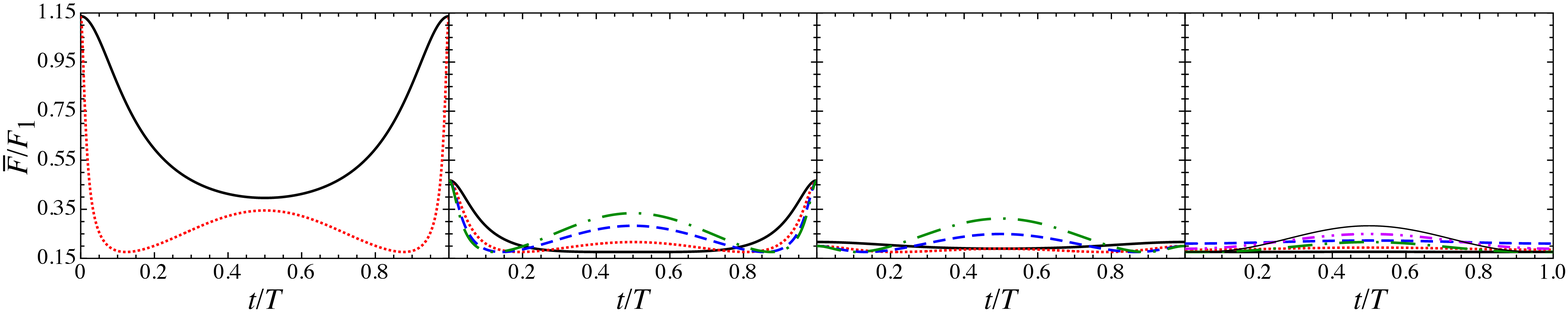}  
\end{tabular}
\end{center}
\caption{
Pulse profiles  for the neutron star with $M=2.21M_\odot$.  Notations are the same as Fig.~\ref{fig:pulse-M200}
}
\label{fig:pulse-M221}
\end{figure*}

Now, we consider the pulse profiles from a rotating neutron star with specific values of $i$ and $\Theta$. The pulse profile with $(i,\Theta)=(a,b)$ is the same as that with $(i,\Theta)=(b,a)$ since Eq.~(\ref{eq:mut_p}) is symmetric with respect to $i$ and $\Theta$. Therefore, we consider the case of $\Theta>i$ in this paper. In particular, we calculate the pulse profiles for the 16 combinations of $i$ and $\Theta$ shown in Fig.~\ref{fig:itheta}, i.e., $\Theta/\pi=0.04$, 0.15, 0.25, 0.35, and 0.45 for $i/\pi=0.02$; $\Theta/\pi=0.15$, 0.25, 0.35, and 0.45 for $i/\pi=0.1$; $\Theta/\pi=0.25$, 0.35, and 0.45 for $i/\pi=0.2$; $\Theta/\pi=0.35$ and 0.45 for $i/\pi=0.3$; $(i/\pi,\Theta/\pi)=(0.4,0.45)$ and $(0.46,0.48)$. These combinations are shown in Fig.~\ref{fig:itheta}, where the left and right panels correspond to the neutron-star models with $M/R=0.2953$ (as an example of $\pi<\psi_{\rm cri}<3\pi/2$) and $M/R=0.3263$ (as an example of $3\pi/2<\psi_{\rm cri}<2\pi$),  respectively. The circles, squares, diamonds, and triangles in the figure correspond to the stellar models for the pulse profiles shown in Figs.~\ref{fig:pulse-M200} and \ref{fig:pulse-M221} from left to right panels.

\begin{figure*}
\begin{center}
\begin{tabular}{ccc}
\includegraphics[scale=0.38]{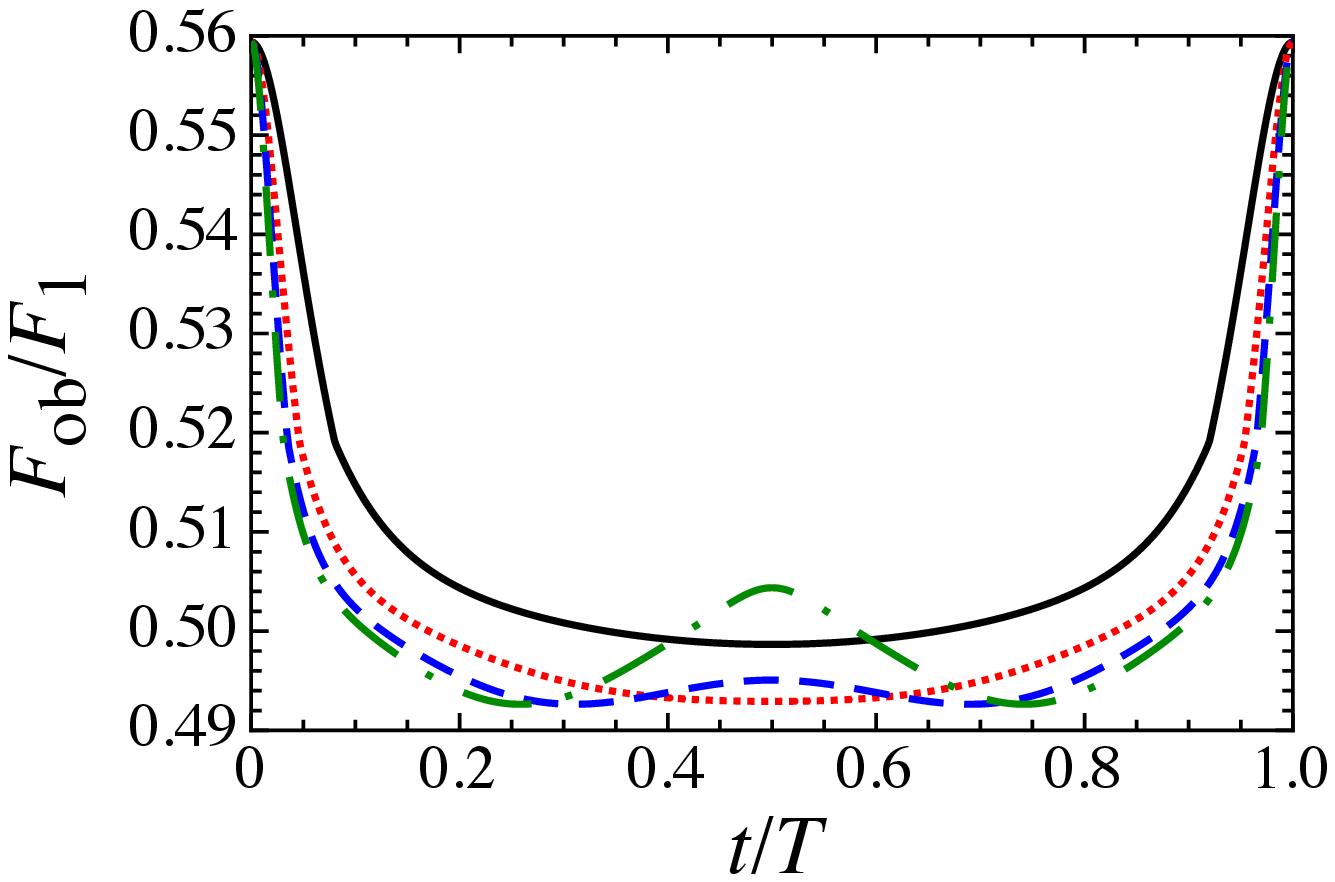} &
\includegraphics[scale=0.38]{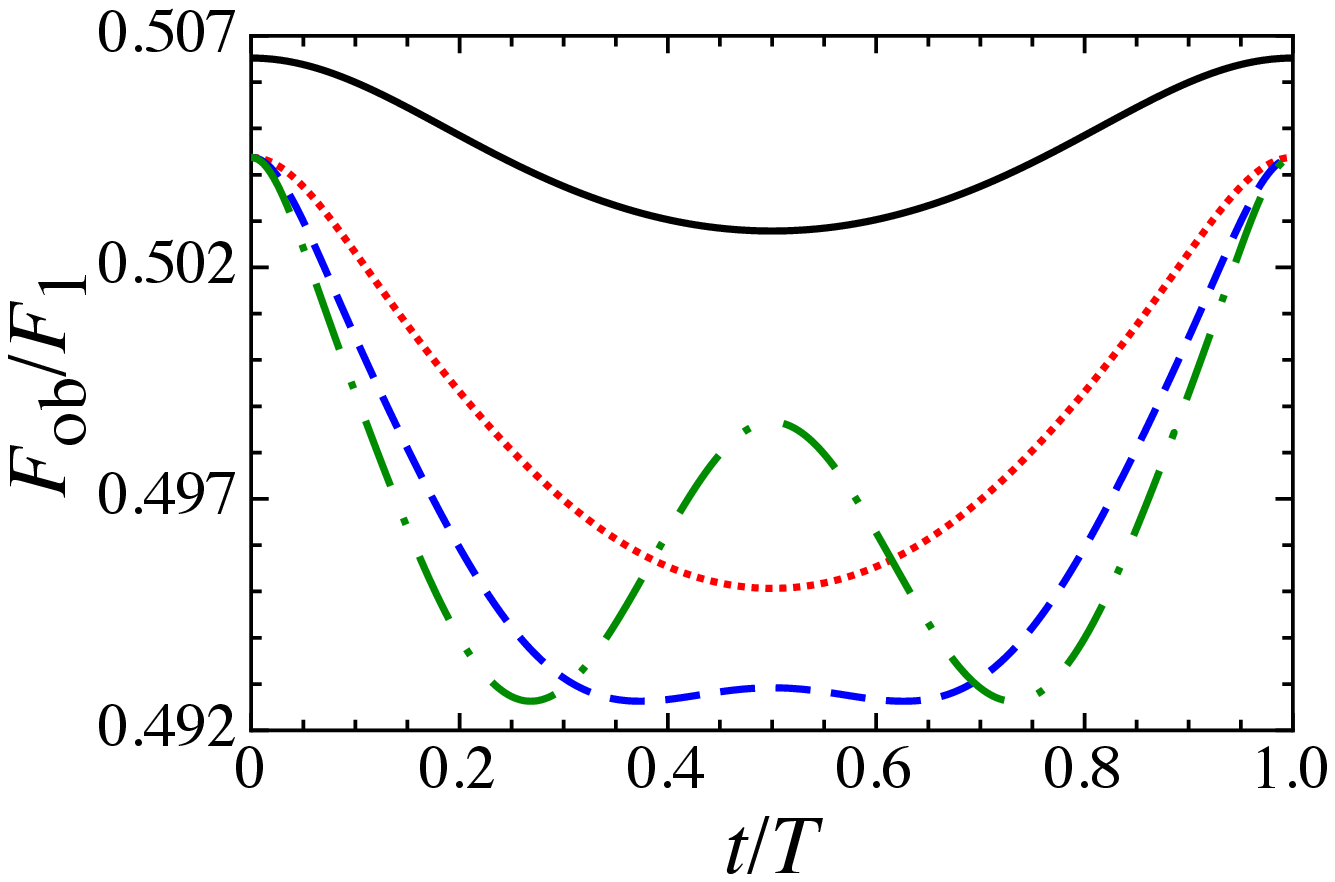} & 
\includegraphics[scale=0.38]{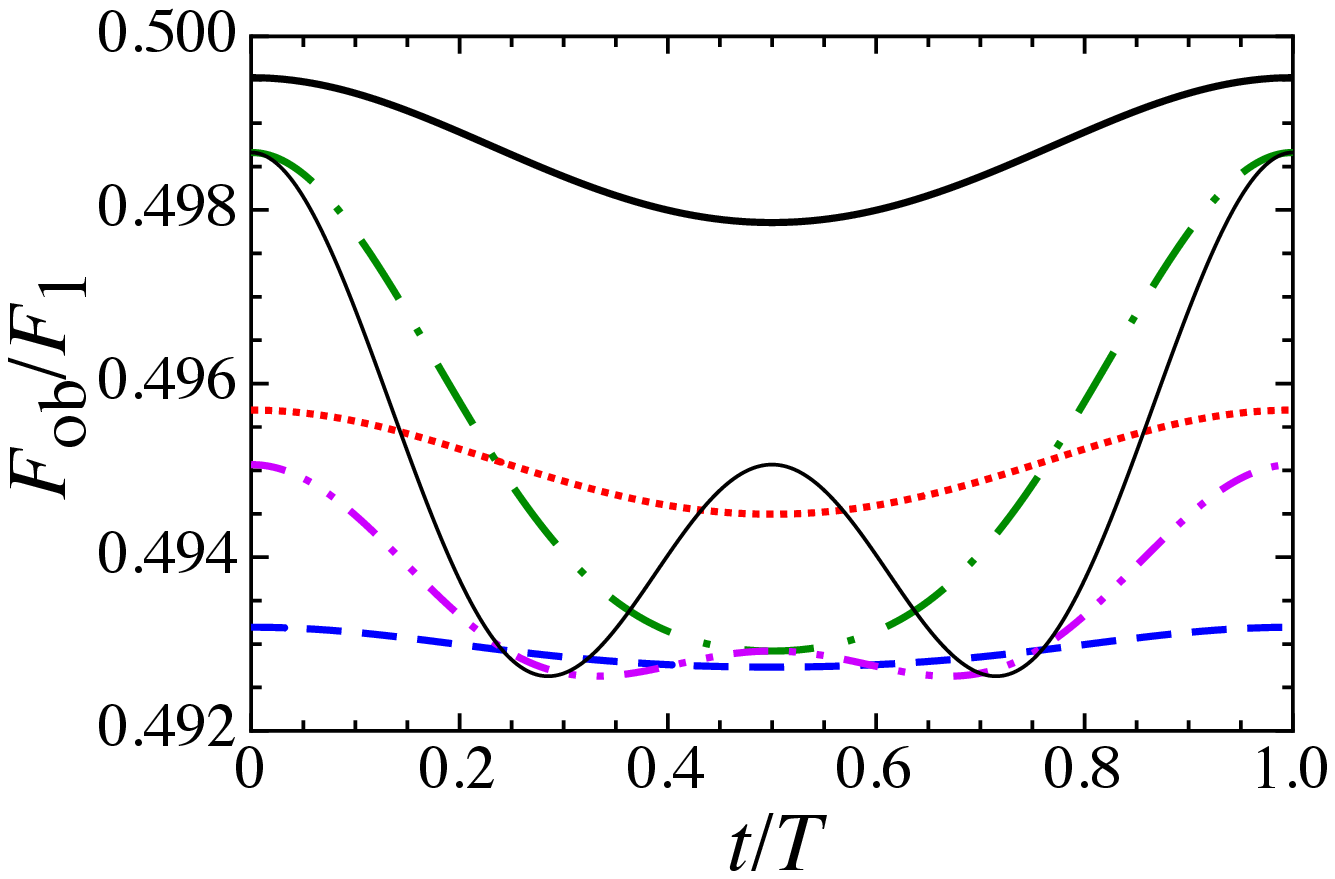}  
\end{tabular}
\end{center}
\caption{
Magnified figures of the observed flux in Fig.~\ref{fig:pulse-M200} except for the leftmost panel. The meaning of lines is the same as the corresponding panels in Fig.~\ref{fig:pulse-M200}.
}
\label{fig:pulse-M200a}
\end{figure*}

In Fig.~\ref{fig:pulse-M200}, the pulse profiles for the neutron-star model with $M/R=0.2953$ are shown as functions of $t/T$ for various combinations of angles $i$ and $\Theta$, where the panels from top to bottom correspond to the observed flux $F_{\rm ob}$, the flux from the primary hot spot $F$, and that from the antipodal hot spot $\bar{F}$, normalized by $F_1$. In the figure, the labels of i, ii, iii, and iv denote the classification shown in Fig.~\ref{fig:theta-i-a0}. The magnified figure of the observed flux except for the leftmost panel in Fig.~\ref{fig:pulse-M200} is presented in Fig.~\ref{fig:pulse-M200a}. As mentioned before, due to the existence of brightening of flux around $\psi\sim\pi$, $F/F_1$ and $\bar{F}/F_1$ does not monotonically vary necessarily during $0\le t/T\le 0.5$. Moreover, we find that the observed flux has maximum at $t=0$ in any cases, i.e., when the primary hot spot is closest to the observer. This feature is different from that in the case of $\psi_{\rm cri}<\pi$, which is presented in Appendix~\ref{sec:a1} (see Fig.~\ref{fig:pulse-M180}).

\begin{figure}
\begin{center}
\includegraphics[scale=0.5]{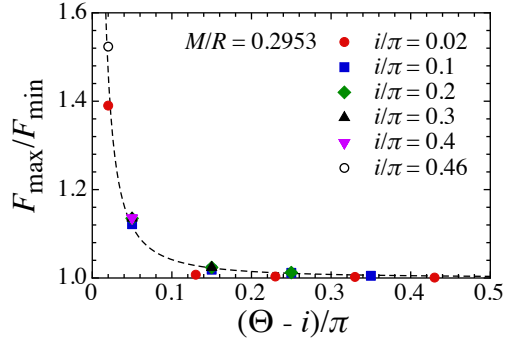} 
\end{center}
\caption{
Ratio of the maximum observed flux $F_{\rm max}$ to the minimum observed flux $F_{\rm min}$ as a function of $(\Theta - i)/\pi$ for the neutron-star model with $M/R=0.2953$. The dashed line is the fitting formula given by Eq.~(\ref{eq:fit-M200}).
}
\label{fig:M200-ratio}
\end{figure}

Furthermore, for the adopted combinations of angles $i$ and $\Theta$, the ratio of the maximum observed flux $F_{\rm max}$ to the minimum observed flux $F_{\rm min}$ for the neutron-star model with $M/R=0.2953$ is shown in Fig.~\ref{fig:M200-ratio} as a function of $(\Theta-i)/\pi$, where the filled-circles, squares, diamonds, triangles, inverted-triangle, and open-circle denote the case for $i/\pi=0.02$, 0.1, 0.2, 0.3, 0.4, and 0.46, respectively. From this figure, one can observe that $F_{\rm max}/F_{\rm min}$ becomes very large as $\Theta-i$ approaches zero. This ratio can be fitted as
\begin{equation}
  F_{\rm max}/F_{\rm min} = \frac{1.311 \times 10^{-3}}{|\Theta - i|^{3/2}}\pi^{3/2} + 1, \label{eq:fit-M200}
\end{equation}
which is also shown in Fig.~\ref{fig:M200-ratio} with dashed line. This is also a different feature compared to the case for $\psi_{\rm cri}<\pi$ (see Fig.~\ref{fig:M180-ratio}), where the ratio of $F_{\rm max}/F_{\rm min}$ depends strongly on the classification of pulse profile as in Fig.~\ref{fig:theta-i-0}.

\begin{figure*}
\begin{center}
\begin{tabular}{ccc}
\includegraphics[scale=0.38]{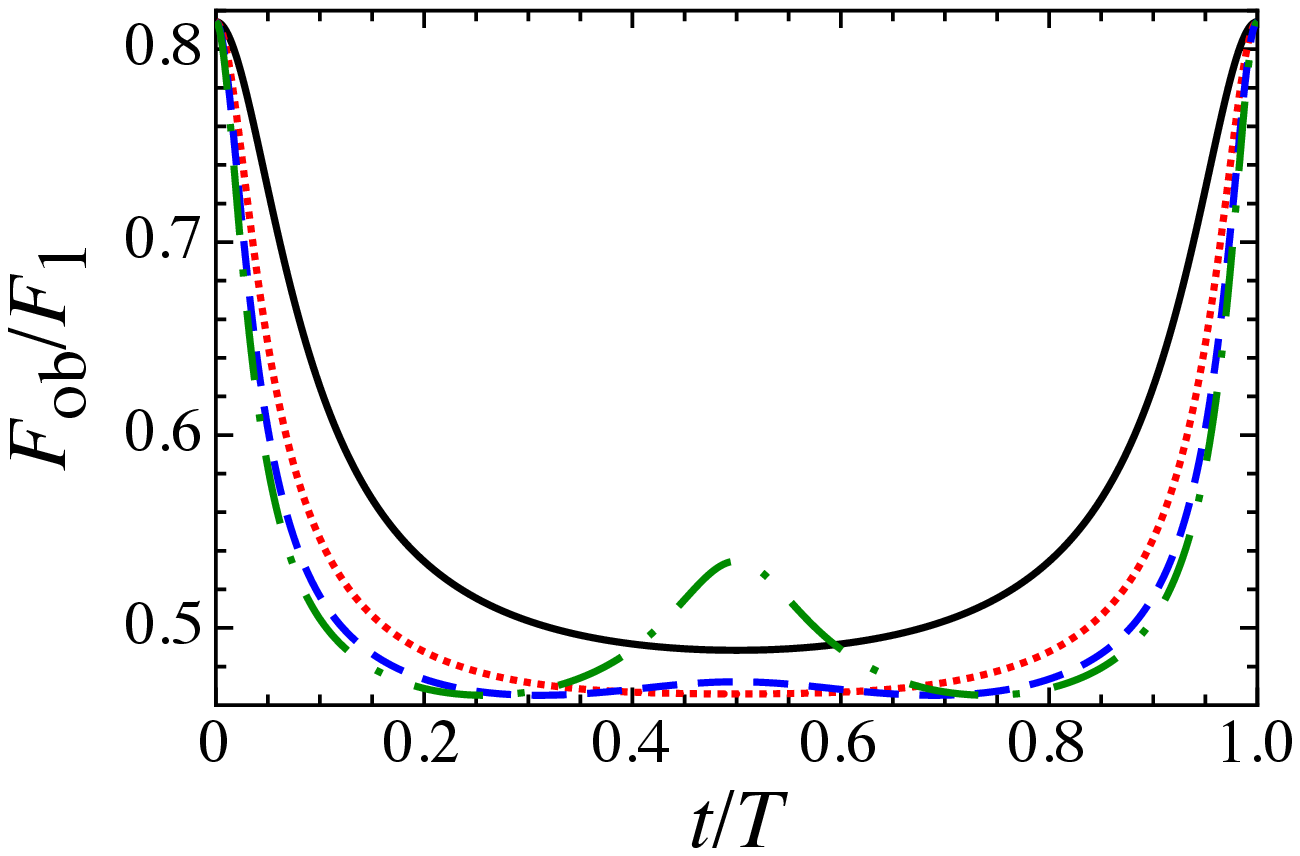} &
\includegraphics[scale=0.38]{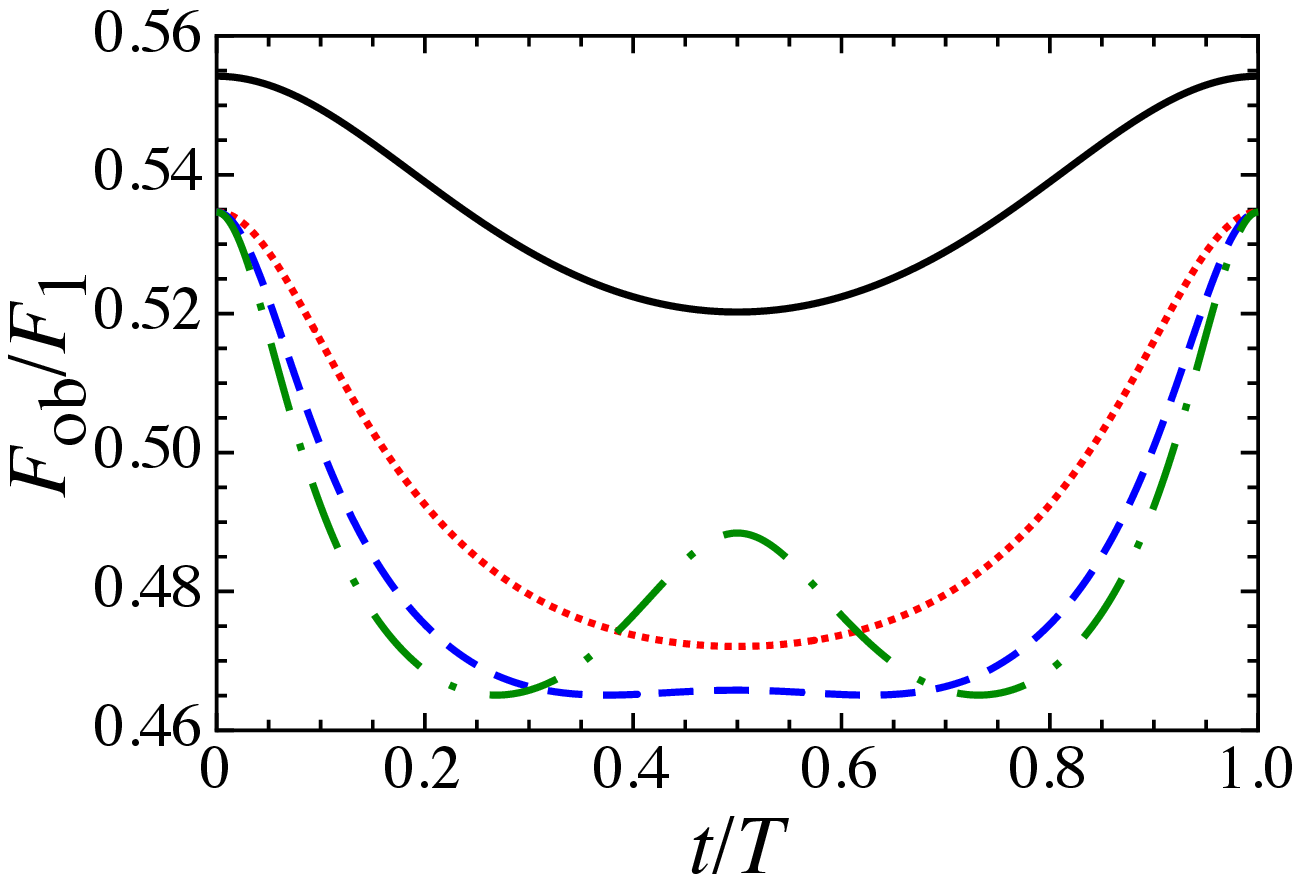} & 
\includegraphics[scale=0.38]{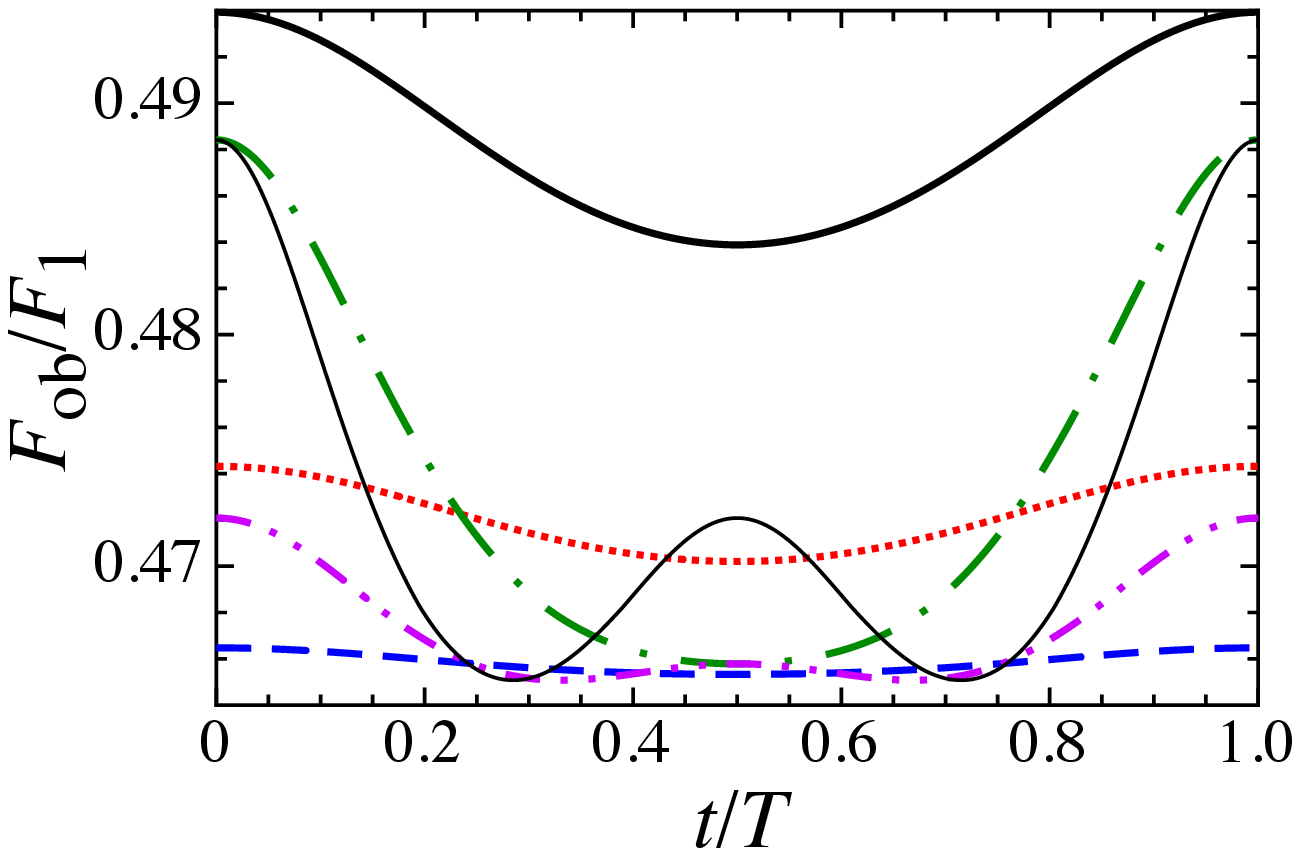}  
\end{tabular}
\end{center}
\caption{
Magnified figure of the observed flux shown in the top panels except for the leftmost panel in Fig.~\ref{fig:pulse-M221}. The meaning of lines is the same as the corresponding panels in Fig.~\ref{fig:pulse-M221}.
}
\label{fig:pulse-M221a}
\end{figure*}

\begin{figure}
\begin{center}
\includegraphics[scale=0.5]{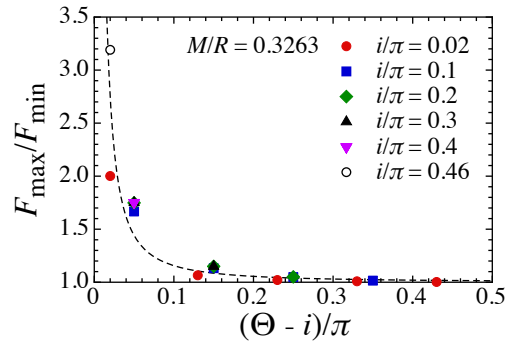} 
\end{center}
\caption{
Ratio of the maximum observed flux $F_{\rm max}$ to the minimum observed flux $F_{\rm min}$ as a function of $(\Theta - i)/\pi$ for the neutron star model with $M/R=0.3263$. The dashed line is the fitting formula given by Eq.~(\ref{eq:fit-M221}). 
}
\label{fig:M221-ratio}
\end{figure}

In a similar way, the pulse profiles from the neutron-star model with $M/R=0.3263$ are shown in Fig.~\ref{fig:pulse-M221}, and the magnified figure of the observed flux in Fig.~\ref{fig:pulse-M221} except for the leftmost panel is presented in Fig.~\ref{fig:pulse-M221a}. In this case, the observed flux is quite similar to that for the neutron-star model with $M/R=0.2953$, although the flux from the primary and antipodal hot spots for $M/R=0.3263$ is different from that for $M/R=0.2953$. Even so, the ratio $F_{\rm max}/F_{\rm min}$ for $M/R=0.3263$ is larger than that for $M/R=0.2953$. Such a ratio is shown in Fig.~\ref{fig:M221-ratio}, of which fitting formula is given by 
\begin{equation}
  F_{\rm max}/F_{\rm min} = \frac{4.935 \times 10^{-3}}{|\Theta - i|^{3/2}}\pi^{3/2} + 1.  \label{eq:fit-M221}
\end{equation}
We remark that the coefficients in Eqs. (\ref{eq:fit-M200}) and (\ref{eq:fit-M221}) should strongly depend on the stellar compactness, because the observed flux depends on the stellar compactness as shown in Fig. \ref{fig:Fpsi}.

\section{Conclusion}
\label{sec:VI}

The light radiating from a compact object is bent due to the strong gravitational field produced by the compact object. As the result, one may observe the photon from the compact object, even if it is radiated from the backside of the object. The bending angle increases as the stellar compactness increases. Even so, most neutron stars are not so compact that the side completely opposite to the observer can be seen, i.e., the invisible zone of star usually exists. However, if the compactness of neutron star is large enough as $M/R\ge 0.2840$, e.g., the neutron star with $M=2.0M_\odot$ and $R=10.40$ km, the invisible zone on the stellar surface disappears. Namely, the photon radiating from any position on the stellar surface can reaches the observer. Since the existence of the $2M_\odot$ neutron stars is known (although the maximum mass of neutron star is not fixed yet), the high compactness $M/R\ge 0.2840$ may be possible (e.g., SLy and APR EOSs). For such a neutron stars, the number of photon paths can be different from that for usual neutron stars. Therefore, in this paper, we made a classification of the number of photon paths, depending on $\Theta$, which is the angle between the rotational axis and the normal vector on the hot spot, and inclination angle $i$.

Then, we calculated the pulse profiles of rotating neutron stars with $M/R=0.2953$ and 0.3263 for the various combinations of angles $i$ and $\Theta$, which were compared with the profiles for the neutron-star model with $M/R=0.2658$ as an example for $M/R\le 0.2840$. As the result, we found that the pulse profiles for $M/R\ge 0.2840$ are qualitatively different from those for $M/R\le 0.2840$. The flux for $M/R\ge 0.2840$ has a maximum when the primary hot spot comes closest to the observer for any $i$ and $\Theta$, while that for $M/R\le 0.2840$ depends strongly on the combination of $i$ and $\Theta$. Additionally, we found that the ratio of the maximum observed flux to the minimum one is significantly large compared to the case for $M/R\le 0.2840$. In particular, such a ratio becomes larger as the stellar compactness increases, and one can see a significant difference between the cases of $M/R\ge 0.2840$ and $M/R\le 0.2840$ in the situation of $|i-\Theta|\sim0$. Thus, the results in this paper suggests that one would be able to constrain the equation of state for neutron stars through the observation of pulse profile, provided angles $i$ and $\Theta$ are determined by other methods.

For simplicity, in this paper we have included the effect of rotation only partially. In fact, as the spin increases up to $\sim$ a few hundred Hz, one will have to take into account the Doppler effect, relativistic aberration~\cite{PG03,PB06}, frame dragging, and the stellar deformation~\cite{CLM05,PO2014}. In addition, for realistic situations, the existence of magnetosphere should be taken into account. Such a situation will be discussed elsewhere.

\acknowledgments
HS is grateful to T. Takiwaki for giving valuable comments. 
This work was supported in part by Grant-in-Aid for Scientific Research (C) through Grant No.\ 17K05458 (HS), 15K05086 (UM), and 18K03652 (UM) provided by JSPS.

\appendix
\section{Pulse profiles for $\psi_{\rm cri} < \pi$}
\label{sec:a1}

\begin{figure}
\begin{center}
\includegraphics[scale=0.5]{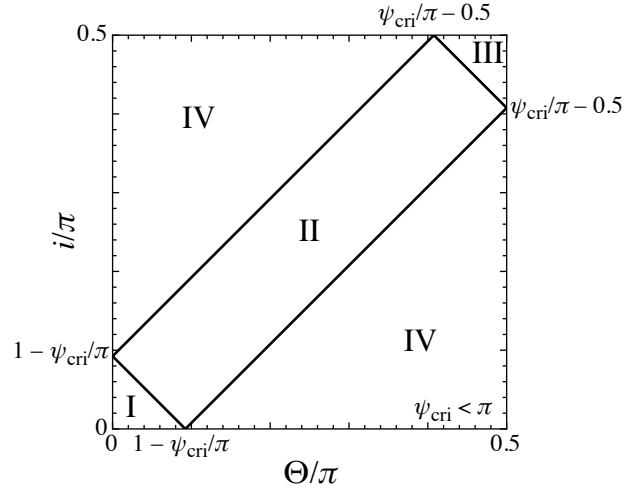} 
\end{center}
\caption{
The plane of $i$ and $\Theta$ is classified depending on whether the primary and antipodal hot spots can be observed or not for the neutron-star model with $M=1.8M_\odot$ and $R=10$ km, for which $\psi_{\rm cri}=0.908\pi$ (i.e., $\psi_{\rm cri}<\pi$)~\cite{SM2017}. The regions denoted by I, II, III, and IV correspond to the cases explained in the text. 
}
\label{fig:theta-i-0}
\end{figure}

For reference, we present the pulse profiles for $\psi_{\rm cri} < \pi$. As mentioned in the text, this situation might be the most likely for ordinary neutron stars, in which the invisible zone exists and its size depends on the stellar compactness. That is, the number of photon paths from the primary and antipodal hot spots is just one. In this situation, whether the primary and antipodal hot spots can be seen is classified into the following four cases, depending on $i$ and $\Theta$~\cite{Beloborodov2002,SM2017}:
\begin{itemize}
  \item[{(I)}]
    Only the primary hot spot is observed at any moment.
  \item[{(II)}]
    The primary hot spot is observed at any moment, while the antipodal hot spot is observed sometime.
 \item[{(III)}]
    The primary hot spot is not observed sometime.
  \item[{(IV)}]
    Both hot spots are observed at any moment. 
\end{itemize}
This classification is depicted in Fig.~\ref{fig:theta-i-0}, where the neutron-star model with $M=1.8M_\odot$ and $R=10$ km is adopted. In order to compare with the case of $\psi_{\rm cri}>\pi$ in the text, we consider the neutron-star models with the specific angles of $i$ and $\Theta$ shown in Fig.~\ref{fig:theta-i-0}.

\begin{figure}
\begin{center}
\includegraphics[scale=0.5]{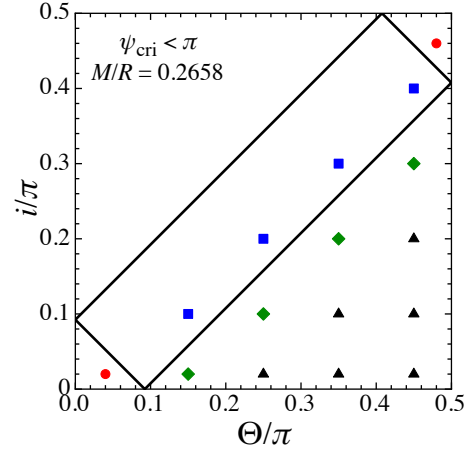} 
\end{center}
\caption{
Specific angles of $i$ and $\Theta$, with which the pulse profiles are computed, and the classification of the plane for the stellar models with $M/R=0.2658$.
}
\label{fig:ithetaM18}
\end{figure}

\begin{figure*}
\begin{center}
\begin{tabular}{c}
\includegraphics[scale=0.38]{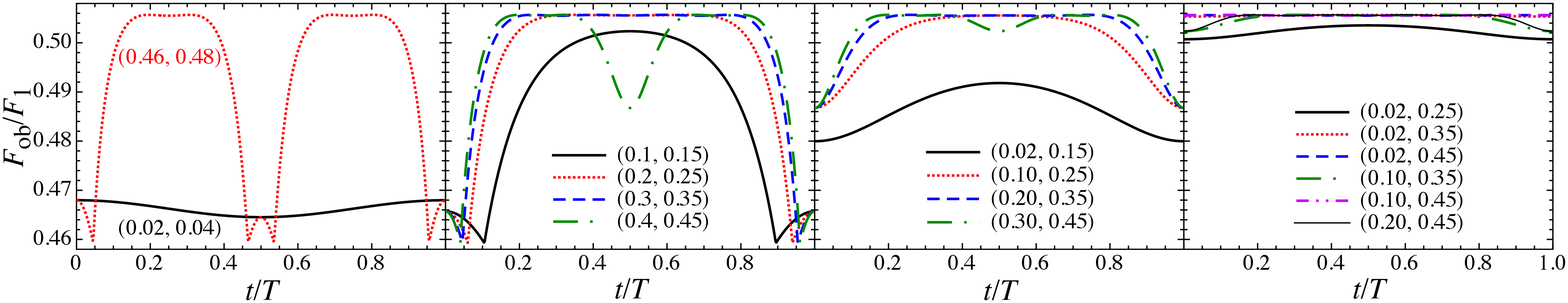} \\
\includegraphics[scale=0.38]{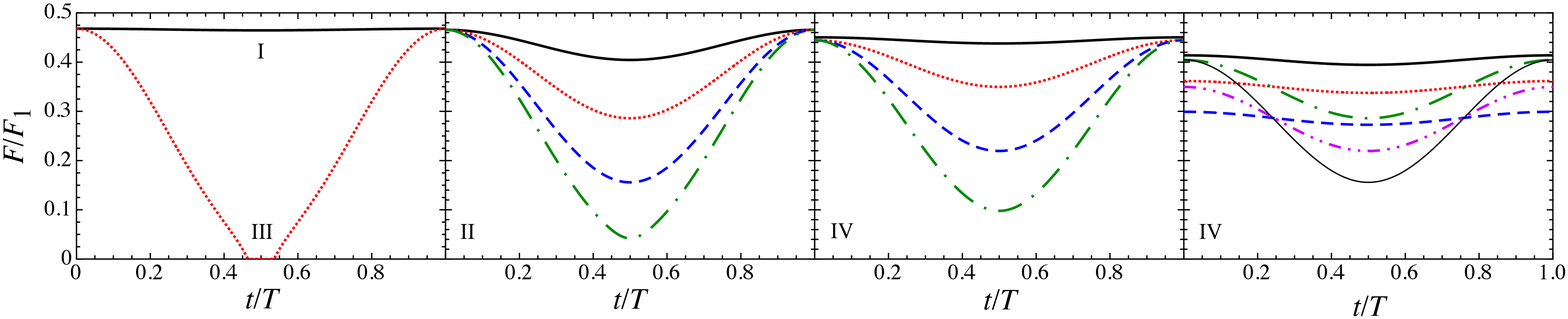} \\ 
\includegraphics[scale=0.38]{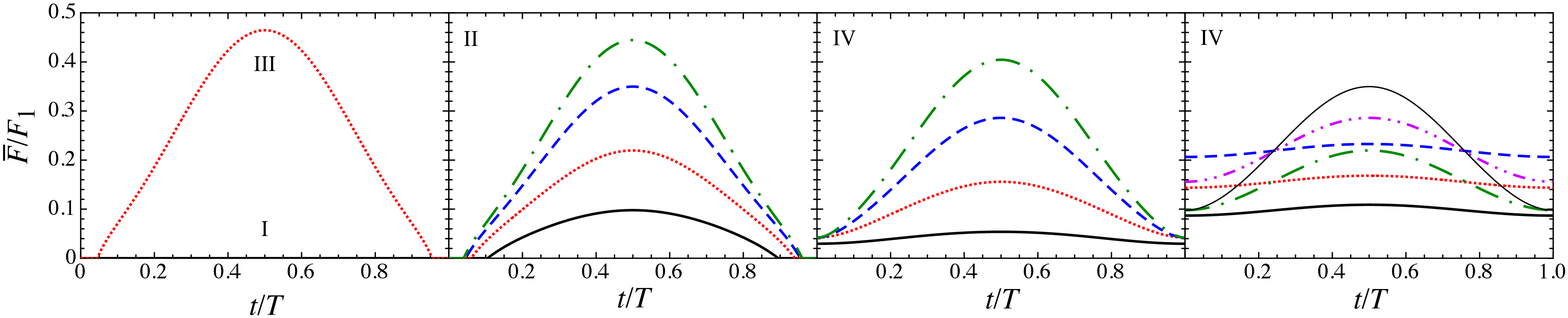}  
\end{tabular}
\end{center}
\caption{
Pulse profiles for the neutron star model with $M=1.8M_\odot$ and $=10$ km are shown as a function of $t/T$ for various angles $i$ and $\Theta$ as shown in Fig.~\ref{fig:ithetaM18}. The panels from left to right correspond to the angles shown in Fig.~\ref{fig:ithetaM18} with the circles, squares, diamonds, and triangles. The panels from top to bottom are the observed flux $F_{\rm ob}$, the flux from the primary hot spot $F$, and the flux from the antipodal hot spot $\bar{F}$, normalized by $F_1$. The labels of I, II, III, and IV denote the class of pulse profiles depending on the angles $i$ and $\Theta$, which is already explained in text. 
}
\label{fig:pulse-M180}
\end{figure*}

\begin{figure}
\begin{center}
\includegraphics[scale=0.5]{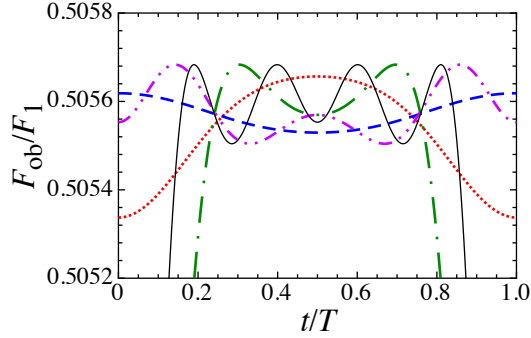} 
\end{center}
\caption{
Magnified figure of the observed flux shown in the top-rightmost panel in Fig.~\ref{fig:pulse-M180}. The meaning of lines is the same as the corresponding panel in Fig.~\ref{fig:pulse-M180}.
}
\label{fig:pulse-M180a}
\end{figure}

\begin{figure}
\begin{center}
\includegraphics[scale=0.5]{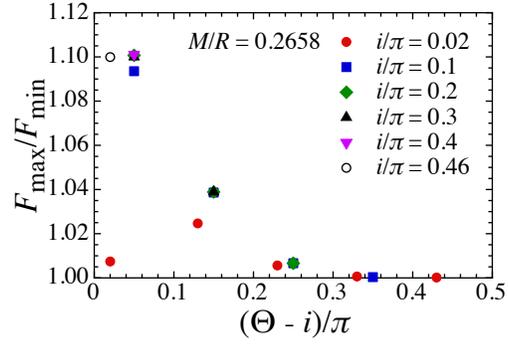} 
\end{center}
\caption{
Ratio of the maximum observed flux $F_{\rm max}$ to the minimum observed flux $F_{\rm min}$ as a function of $(\Theta - i)/\pi$ for the neutron star model with $M/R=0.2658$.
}
\label{fig:M180-ratio}
\end{figure}

We present the combinations of angles $i$ and $\Theta$ in Fig.~\ref{fig:ithetaM18} together with the boundary of classification of pulse profiles, which is the same as in Fig.~\ref{fig:itheta}. The pulse profiles for $M/R=0.2658$ are calculated for these combinations of angles. The obtained pulse profiles are shown in Fig.~\ref{fig:pulse-M180}, and the magnified figures of the observed flux is shown in Fig.~\ref{fig:pulse-M180a}. Unlike the case of $\psi_{\rm cri}>\pi$ discussed in the text, the time at which the observed flux has a maximum depends on the class of pulse profiles, i.e., the combination of $i$ and $\Theta$. Additionally, one can see that the flux from the primary (resp.\ antipodal) hot spot monotonically decreases (rest.\ increases) during $0\le t/T\le0.5$ as expected from Fig.~\ref{fig:Fpsi}, which is a different behavior from that in the case of $\psi_{\rm cri}>\pi$.  Furthermore, the ratio of $F_{\rm max}$ to $F_{\rm min}$ is presented in Fig.~\ref{fig:M180-ratio} as a function on $(\Theta-i)/\pi$. Again, the behavior for $\psi_{\rm cri}<\pi$ is different from that for $\psi_{\rm cir}>\pi$. Namely, the ratio depends on the class of pulse profiles, and one cannot say that the ratio increases as $|\Theta-i|/\pi$ decreases. In addition, one can observe that the ratio for $\psi_{\rm cri}<\pi$ is significantly smaller than that for $\psi_{\rm cir}>\pi$.

\end{document}